\documentclass[a4paper,fleqn]{cas-dc}

\usepackage[numbers]{natbib}
\usepackage{rotating}

\begin{document}
\let\WriteBookmarks\relax
\def\floatpagepagefraction{1}
\def\textpagefraction{.001}
\shorttitle{Numerical optimization of aviation decarbonization scenarios}
\shortauthors{I. Costa-Alves et~al.}

\title [mode = title]{Numerical optimization of aviation decarbonization scenarios: balancing traffic and emissions with maturing energy carriers and aircraft technology}



\author[1,2]{Ian Costa-Alves}[type=editor,
                        auid=000,bioid=1,
                        orcid=0009-0009-4458-6641]
\cormark[1]
\fnmark[1]
\ead{ian.costa-alves@isae-supaero.fr}

\credit{Conceptualization of this study, Methodology, Software}

\affiliation[1]{organization={Aerodynamics, Energetics and Propulsion Department, ISAE-SUPAERO},
                addressline={10 av. Marc Pélegrin}, 
                city={Toulouse},
                postcode={31055}, 
                country={France}}

\affiliation[2]{organization={Multidisciplinary Optimization Competence Center, IRT Saint Exupéry},
                addressline={3 rue Tarfaya}, 
                postcode={31400}, 
                city={Toulouse},
                country={France}}

\author[1]{Nicolas Gourdain}
\ead{nicolas.gourdain@isae-supaero.fr}

\author[2]{François Gallard}
\ead{francois.gallard@irt-saintexupery.com}


\credit{Data curation, Writing - Original draft preparation}

\author[2]{Anne Gazaix}
\ead{anne.gazaix@irt-saintexupery.com}

\author[1,3]{Yri Amandine Kambiri}

\author[3]{Thierry Druot}

\affiliation[3]{organization={Conceptual Airplane Design and Operations, ENAC},
                addressline={7 av. Marc Pélegrin}, 
                city={Toulouse},
                postcode={31400}, 
                country={France}}



\begin{abstract}
Despite being considered a hard-to-abate sector, aviation’s emissions will play an important role in long-term climate mitigation of transportation. The introduction of low-carbon energy carriers and the deployment of new aircraft in the current fleet are modeled as technology-centered decarbonization policies, while supply constraints in targeted market segments are modeled as demand-side policies. Shared Socioeconomic Pathways (SSPs) are used to estimate trend-mitigation traffic demand and to limit the sectoral consumption of electricity and biomass. Mitigation scenarios are formulated as optimization problems, and three applications are demonstrated: no-policy baselines, single-policy optimization, and scenario-robust policies. Results show that the choice of energy carrier is highly dependent on assumptions regarding aircraft technology and the background energy system. Across all SSP-based scenarios, emissions peak by around 2040, but achieving alignment with the Paris Agreement requires either targeted demand management or additional low-carbon energy supply. The use of gradient-based optimization within a multidisciplinary framework enables the efficient resolution of these nonlinear, high-dimensional problems while reducing implementation effort.
\end{abstract}

\begin{graphicalabstract}
\includegraphics[width=\textwidth]{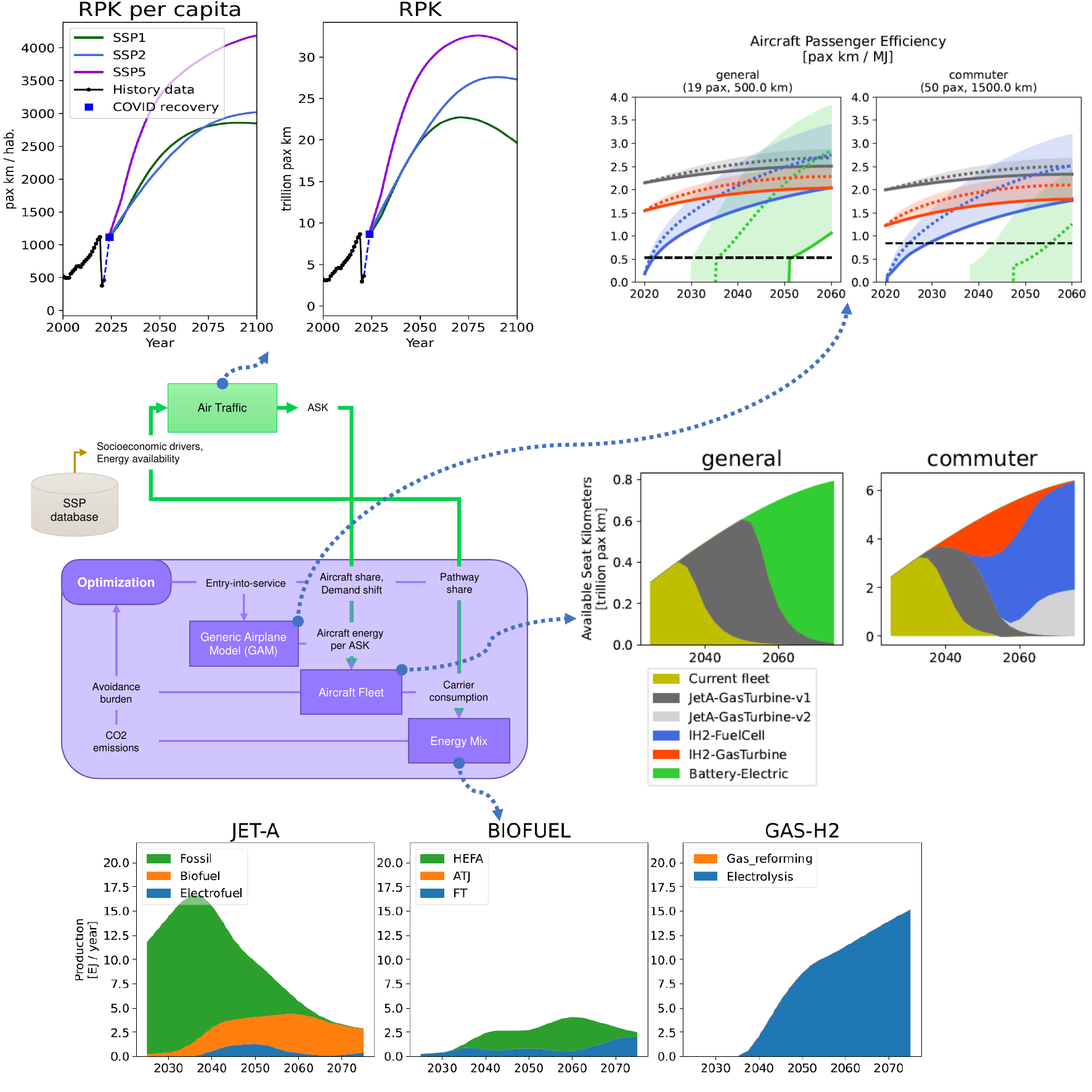}
\end{graphicalabstract}

\begin{highlights}
\item Optimization framework links SSP scenarios with detailed aircraft technology;
\item Fleet replacement and demand saturation leads to emissions peak around 2035 to 2040;
\item Impact of alternative aircraft requires extending analysis beyond 2050;
\item Respecting +2°C carbon budgets requires demand caps or extra energy availability;
\item Policy optimization was prohibitive without speedups from numerical methodology.
\end{highlights}

\begin{keywords}
Multidisciplinary Optimization \sep Low-carbon fuels \sep Aircraft design \sep Integrated Assessment Models \sep Shared Socioeconomic Pathways
\end{keywords}

\maketitle

\section{Introduction}
\label{sec:intro}

Air transportation is often considered a hard-to-abate sector, because carbon-free commercial aircraft are not readily available, and the production of alternative energy carriers to fossil kerosene on a global scale requires significant amounts of biomass and low-carbon electricity to meet the growing demand \cite{becken_implications_2023}. However, the sector will play a critical role in long-term climate mitigation of transportation \cite{girod_global_2012}.

Several key energy carriers have emerged as potential substitutes of conventional kerosene, such as synthetic kerosene (from biomass-to-liquid or power-to-liquid pathways), liquid hydrogen, ammonia, liquid natural gas, ethanol, methanol, and batteries \cite{ansell_review_2023}. Among these, only synthetic kerosene can be used in today’s fleet without aircraft re-deisgn. Current engines are limited by certification to a 50 \% mixing ratio with conventional kerosene, but there is ongoing advances on the operation with 100 \% Sustainable Aviation Fuels (SAF) \cite{full-saf}. There is, however, among industry roadmaps a systematic over-reliance on SAF deployment, which could require 9 \% of global renewable electricity and 30 \% of sustainably available biomass by 2050 \cite{becken_implications_2023}.

Compared to ground and marine transportation vehicles, aircraft are more weight sensitive because they must generate lift in order to remain airborne. Increasing mass leads to an increase in induced drag, which further increases the structural mass to support extra loads, leading to more drag, and so on. This snowball effect is a central phenomenon in the conception of aircrafts \cite{raymer_aircraft_2018}, which means that estimating energy consumption is a fundamentally coupled problem. Alternative aircraft designs, flying with different energy carriers, may display higher energy consumption and/or limited payload and ranges compared to conventional jet-fueled planes, due to low specific energy (batteries) or due to the need for heavier cryogenic fuel tanks (liquid hydrogen).


\begin{quote}
    "Since the IPCC’s Fifth Assessment Report (AR5) there has been a growing awareness of the need for demand management solutions combined with new technologies, such as the rapidly growing use of electromobility for land transport and the emerging options in advanced biofuels and hydrogen-based fuels for shipping and aviation." \cite[][Executive Summary]{transport_ar6_wg3}
\end{quote}

Integrated Assessment Models (IAM) are the tools used to simulate the evolution of the coupled climate-energy-land system from socioeconomic assumptions on climate mitigation and adaptation. Since the publication of the Shared Socioeconomic Pathways (SSP) \cite{riahi_shared_2017}, great improvements have been made in the detail given to transportation modes within global mitigation. The modeling of aviation within IAM's, has shown how demand-side measures may play an important role in keeping temperature increase between 1.5 and 2° C above preindustrial levels \cite{sharmina_decarbonising_2021}, but also how the incorporation of synthetic kerosene and the deployment of new aircraft with alternative energy carriers (such as hydrogen and biofuels) can also play a significant role in reaching stringent targets \cite{napp_role_2019, speizer_integrated_2024}.

However, the detail these studies have concerning the fleet composition and how they are arranged within market segments is limited, e.g., \cite{speizer_integrated_2024} considers kerosene, electric, and hydrogen aircraft to have similar energy consumption even for long-haul (where electric is unfeasible, and hydrogen is less efficient), while \cite{napp_role_2019} accounts for lower efficiencies for hydrogen, it does so with a singular value applied for the entire fleet. The performance of hydrogen aircraft is highly dependent on their range and the cryogenic tank technology \cite{adler_hydrogen-powered_2023}, which can lead to increased energy consumption \cite{icct_hydrogen} or limited operation in terms of payload and ranges \cite{icct_fuelcell}. For electric aircraft, this is even more pronounced as the battery weight significantly shortens the maximum achievable range and payload, making these aircraft only suited to regional flights \cite{icct_electric}.

Sectoral specific scenarios can account for this either with a detailed network of Origin-Destination pairs \cite{grewe_evaluating_2021, eaton_regional_2024, hoelzen_h2-powered_2025}, or by separating the traffic into market segments according to flown distance. A review of sectoral aviation scenarios \cite{delbecq_sustainable_2023} shows that methodologies may differ regarding: traffic evolution, mitigation levers, inclusion of non-CO2 effects, and resource consumption. Employing open-source tools, such as the AeroMAPS framework \cite{planes_aeromaps_2023}, to simulate these types of scenarios can be greatly beneficial to explicit modeling assumptions and finding a common ground for high-level decision making.

From a technological perspective, it is important to map the range of components that are required for novel system architectures. In the context where many of such components are still at immature technologies and require significant allocation of resources to develop, quantifying upper and lower bounds associated to each component is paramount to compare architectural choices in terms of system-level performances \cite{kirby_forecasting_1999} and in end-point metrics, such as financial returns \cite{de_weck_technology_2022}. Regarding climate mitigation, determining the best "carbon-neutral" propulsion architecture choice for each flight is subject of recent interest \cite{adler_energy_2025}, but rely on the strong assumption of dedicated renewable electricity to power aviation, which may be supply constrained.

\subsection{Research positioning}

Global IAM's lack technology and fleet detail concerning aviation emissions, while industry roadmaps and technology assessments rely on strong assumptions of abundant renewable energy. We propose to bridge these together using optimization to endogenously chose appropriate timing and market penetration of conventional and alternative aircraft concepts, subject to a background scenario using the AR6 scenario database \cite{ar6_database}.

These policy scenarios are nonlinear, with many variables due to disaggregation (per market and technologies), and time-dependent (high-dimensional arrays). In practice, even the simpler policy optimization problems may: fail to converge under gradient-free nonlinear optimizers, display significant online linearization cost or offline implementation burden for gradient-based optimization. The numerical methodology aspect of this paper is therefore a central contribution, reducing policy optimization time from 1.5 hours to a single minute on a conventional laptop.

This work builds on the literature of aviation climate mitigation scenarios by linking sectoral mitigation with a background system based on the Shared Socioeconomic Pathways (SSP). Socioeconomic drivers are used to estimate future traffic volumes, based on the assumption that historical trends on traffic growth from personal income and population growth remain unchanged. Biomass and electricity consumed by scenarios are limited by using the concept of a fair share of global production that is allocated to the sector, which avoids over-consumption that can be detrimental to other economic sectors. Fleet and technology detail are also incorporated by modeling conceptual design and possible Entry-Into-Service (EIS) of 4 aircraft propulsion architectures, each linked to component-level performances that mature over time.

This methodology was introduced in \cite{costaalves-isabe}, which addressed the exploration of technology-based decarbonization scenarios with optimization algorithms. Novel developments include the incorporation of voluntary and price-based demand avoidance strategies, the timing of introduction of new technologies, and scenario-robust optimizations.

It is important to note that the generated scenarios are considered as decarbonization scenarios rather than climate mitigation scenarios, because non-CO2 emissions are unaccounted for. While these effects can make up for more than half of aviation's contribution to global warming in the 2000-2018 period \cite{lee_contribution_2021}, there are still significant uncertainties concerning their estimation, especially when considering novel propulsion systems, for which little data is available. Furthermore, because of the short-lived nature of these warming effects, the debate on which metrics to use for comparing them to CO2 is still ongoing \cite{megill_alternative_2024}. However, some findings indicate that designing aircraft to fly lower and slower is capable of significantly reducing non-CO2 effects, with little extra fuel and cost penalties \cite{proesmans_thesis_2024}, these are incorporated in the aircraft design requirements (subsection \ref{subsec:aircraft-design}).

While the usage of optimization within IAM's is not new \cite{nordhaus_optimal_1992, barrage_policies_2023, gcam-2019, witness}, there are several limitations with the tools used to solve them, which often imply a simplification of the problem: reducing time resolution, linearization of the problem \cite{huppmann_messageix_2019}, or even simplification of two-way couplings \cite{jgcricassandra_2024}. Multidisciplinary Optimization (MDO) frameworks can offer methods that allow for: formalizing the coupled problem within optimization routines, and using gradient-based algorithms, which allows nonlinear optimization to scale well despite increasing number of variables. On this methodological front, the present work also contributes to the formulation of optimal mitigation policies, and to alleviate the main drawback of using IAM's with MDO: the extra implementation burden, which is dealt with by using libraries with automatic differentiation capabilities.

The paper is organized as follows: first the quantitative models used are described and compared with the state-of-the-art, secondly the methodology behind optimal scenario formulation and implementation are discussed and initial computational gains are presented, then results of elementary models and optimal mitigation scenarios are presented and analyzed, the shortcomings of models are made explicit in the limitations section, and a critical view on the results and implications from the modeled future are addressed in the discussion, and finally the conclusion on the contributions this research brings are made.

\section{Methods}
\label{sec:methods}

\begin{figure*}
	\centering
	\includegraphics[width=\textwidth]{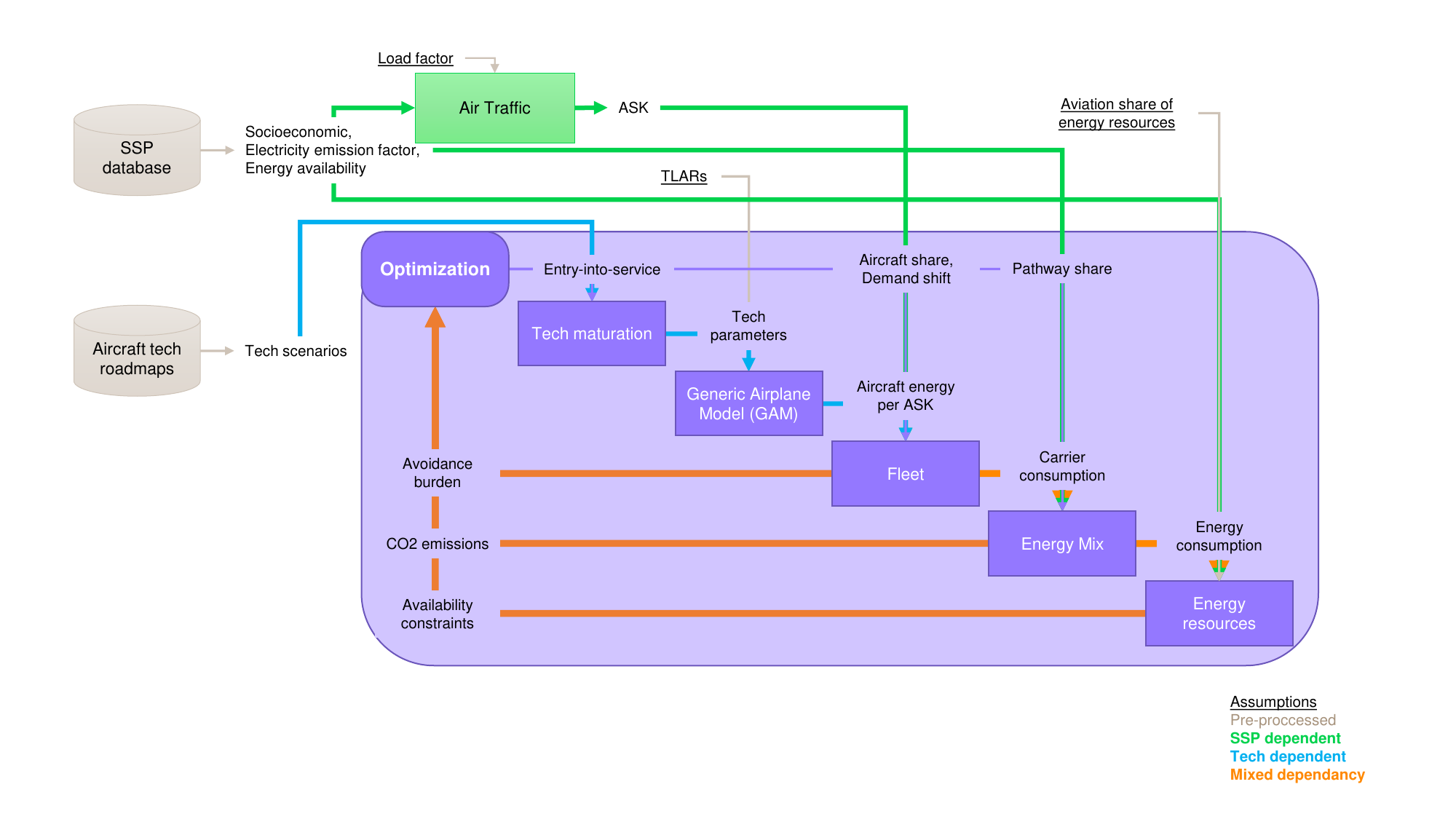}
	\caption{Conceptual view of the optimization process and data-flow between modeled disciplines.}
	\label{fig:models}
\end{figure*}

An overview of the data-flow between the model disciplines is presented in Figure \ref{fig:models}. Before the optimization loop starts, the chosen global scenario determines the evolution of socioeconomic drivers (population and economy) and energy system (global production of biomass and electricity, and emission factor of grid electricity). Then, given the optimization loop varies the set of policy variables, based on simulation objective and constraints.

\subsection{Background scenarios}

The Shared Socioeconomic Pathways (SSP) were introduced between IPCC's 5th and 6th Assessment Report (AR5 and AR6) as a way to bridge the socioeconomic dimension of mitigation into climate assessments \cite{riahi_shared_2017}. These scenarios are formulated based on qualitative stories first, which define high-level assumptions that steer IAM simulations into different possible futures regarding economic development, demographics, energy production, and land-use. Each SSP is created from a storyline to represent a consistent underlying logic to the depth of socioeconomic changes that societies are expected to have:

\begin{itemize}
    \item SSP1: Sustainability – Taking the Green Road (Low challenges to mitigation and adaptation): “The world shifts gradually, but pervasively, toward a more sustainable path, emphasizing more inclusive development that respects perceived environmental boundaries. Driven by an increasing commitment to achieving development goals, inequality is reduced both across and within countries. Consumption is oriented toward low material growth and lower resource and energy intensity” \cite{ssp1}.
    \item SSP2: Middle of the Road (Medium challenges to mitigation and adaptation): “The world follows a path in which social, economic, and technological trends do not shift markedly from historical patterns. Development and income growth proceeds unevenly, with some countries making relatively good progress while others fall short of expectations. Global and national institutions work toward but make slow progress in achieving sustainable development goals” \cite{ssp2}.
    \item SSP5 Fossil-fueled Development – Taking the Highway (High challenges to mitigation, low challenges to adaptation): “This world places increasing faith in competitive markets, innovation and participatory societies to produce rapid technological progress and development of human capital as the path to sustainable development. Global markets are increasingly integrated. There are also strong investments in health, education, and institutions to enhance human and social capital. At the same time, the push for economic and social development is coupled with the exploitation of abundant fossil fuel resources and the adoption of resource and energy-intensive lifestyles around the world” \cite{ssp5}.
\end{itemize}

After the adoption of the Paris Agreement, studies focused on the investigation of pathways limiting warming to 1.5°C \cite{ipcc_wg3_an3} in the context of the Special Report on Global Warming of 1.5°C (SR1.5). From then until AR6, significant work has been done in improving the modeling of energy and transportation technologies, and in multi-model studies covering scenarios from extrapolation current policy trends and the implementation of NDCs, all of them are made available in the AR6 scenario database \cite{ar6_database}.

From each baseline, mitigation scenarios are introduced by incorporating varying mitigation strategies and are named according to a target radiative forcing by the end of the century, these are made to match scenario emissions to a Representative Concentration Pathway (RCP), which are used in the analysis carried by the IPCC Working Groups 1 and 2. These scenarios are then incorporated by main modeling groups, each resposible for one IAM, forming an SSP-RCP-IAM matrix (see \cite{ipcc_wg3_an3}).

The choice of the background scenario determines the background population and GDP of the entire economy, the total energy production for biomass and grid electricity (considered further as primary energy inputs to the aviation energy system), as well as the associated electricity emission intensity. The simulated results, are based on scenarios: SSP1 with RCP 1.9 from model WITCH-GLOBIOM 3.1, SSP2 with RCP's 1.9, 2.6, and 3.4 from MESSAGE-GLOBIOM 1.0, and SSP5 with RCP 4.5 from REMIND-MAgPIE 1.5. Then in the validation of the final results, the aviation sector's final energy and emissions are compared with the entire scenario ensemble from models IMAGE 3.2, REMIND-MAgPIE 2.1-4.3, REMIND-Transport 2.1, and AIM/Hub-Global 2.0.

\subsection{Models}

An overview of key models behind scenario generation is provided here, for a more detailed explanation on the model assumptions and equations refer to the Annex (section \ref{sec:models}). In early studies \cite{costaalves-isabe}, models primarily aimed to link SSP scenario data and aircraft design routines within an aviation system model, through a direct re-implementation of AeroMAPS \cite{planes_aeromaps_2023}. The models and numerical methods developed in the present work will later be incorporated into AeroMAPS.

First, trend scenarios of air traffic demand, expressed in terms of Revenue Passenger Kilometers (RPK), are generated from socioeconomic drivers linked to a global scenario in the AR6 database \cite{ar6_database}. The air traffic supply, in terms of Available Seat Kilometers (ASK), is then estimated using a re-implementation of AeroMAPS \cite{planes_simulation_2021, planes_aeromaps_2023} load factor model.

Aircraft energy consumption is estimated using the Generic Airplane Model (GAM) model for a set of propulsion architectures and market segments, accounting for technology available by entry-into-service (EIS), given a set of fixed Top Level Aircraft Requirements (TLAR) per aircraft and per market. Covered propulsion systems in this study are:
\begin{itemize}
    \item Gas turbines powered by Jet-A
    \item Gas turbines powered by liquid hydrogen (lH2)
    \item Electric propulsion powered by batteries
    \item Electric propulsion powered by liquid hydrogen and fuel cells
\end{itemize}

The final consumption of energy carriers are estimated per each market, accounting the operation of a bottom-up aircraft fleet, and a part of trend demand that is not met due to demand avoidance policy.

There are many types of produced energies, some energy types assemble other energies in a mixing process (such as Jet-A fuel, encompassing fossil fuel and SAF blends) each may be produced with a set production pathways, each of them consuming other energy types as inputs (see Table \ref{tab:energy-prod}):
\begin{itemize}
    \item Jet-A fuel: Blended mix of fossil kerosene, biofuel, and electrofuel
    \item Fossil kerosene: Refineries, consume oil and have direct emissions
    \item Biofuel: Biomass-to-Liquid (BtL), consume biomass and have direct emissions
        \begin{itemize}
            \item Hydroprocessed Esters and Fatty Acids (HEFA)
            \item Alcohol-to-Jet (ATJ)
            \item Fischer-Tropsch (FT)
        \end{itemize}
    \item Electrofuel: Power-to-Liquid (PtL), consumes electricity and gas H2
    \item Liquid H2: Hydrogen liquefaction, consumes electricity and gas H2
    \item Gas H2:
    \begin{itemize}
        \item Gas reforming, has direct emissions
        \item Electrolysis, consumes electricity
    \end{itemize}
    \item Batteries: Charging, consumes electricity
\end{itemize}

The energy inputs consumed to meet the necessary final energy carrier production and the emissions generated are estimated from process efficiencies, direct emissions from pathway production and grid electricity emission factor linked to a global scenario. In the present model, each biofuel pathway competes for the same biomass feedstock, considering direct emissions and process efficiencies as an average of the different feedstock-specific processes \cite{NEULING201854}. However, this approach represents a simplified view, since each biofuel pathway can actually consume specific feedstocks, such as energy crops (competing with food production), oil-based feedstocks and wastes. For example, HEFA biofuels can be obtained from oil-based feedstocks (such as jatropha and camelina), but also from used cooking oil. The final climatic impact of biofuels also depends on potential land use change, especially in the case of competition with food production \cite{yang_sustainable_2025}.

Finally, the consumption of grid electricity and biomass are constrained using the concept of an allocated share of global production of energy resources (biomass and electricity). As the choice of allocated resources dedicated to aviation is ultimately an outcome of regulations and market mechanisms, two different values are considered to explore the consequences of trend and extra energy availability.

The modeling of costs (energy, aircraft, operations) is outside the scope of this paper. Under the current production system, the price of alternative energy carriers is higher than that of fossil kerosene \cite{salgas_cost_2023, salgas_marginal_2024}, but this gap is expected to decrease in the near future due to scaling of production and due to carbon pricing \cite{dray_cost_2022}, especially in scenarios with ambitious climate targets.

\subsubsection{Air Traffic Demand}

\begin{figure*}
    \centering
    \includegraphics[width=\textwidth]{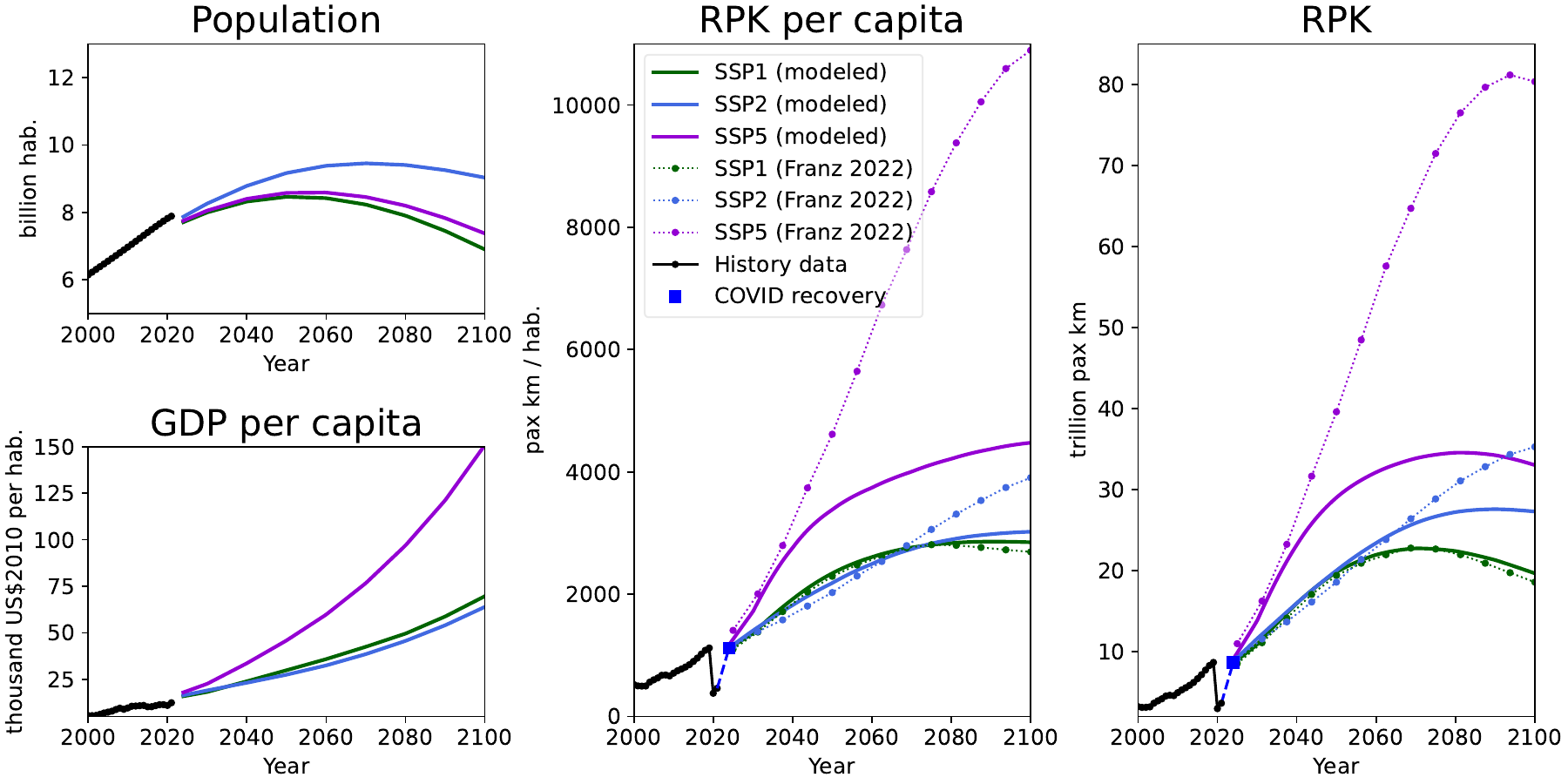}
    \caption{Historical evolution and modeled future global RPK demand from socioeconomic drivers linked to SSP scenarios. Results are compared with \cite{franz_wide_2022}. Explicitly incorporating per-capita demand saturation (current model) yields much lower traffic levels in the long term, when compared with models based on the extrapolation of short term income elasticities.}
    \label{fig:global-demand}
\end{figure*}

In order to account for the phases of growth, maturation and saturation observed in aviation demand \cite{vedantham_aircraft_1994}, a generalized logistic function is used to estimate trend per-capita demand from per-capita income. This model explicitly accounts for saturation of per-capita transport demand as per-capita incomes grow, differently to the approach in most IAMs, which rely on extrapolations \cite{franz_wide_2022} or assumptions of constant income elasticity \cite{kim_gcam_2006, gcam-2019}.

Figure \ref{fig:global-demand} shows the recent history and SSP scenario future evolution of Population and GDP per capita, and also the recent history and modeled future global RPK demand. SSP scenario 2 follows the historically calibrated logistic function, demand grows with varying pace due to income growth, but stabilizes at around 3000 passenger-kilometers per capita. For SSP1 displays high initial growth due to income, but stabilizes demand sooner and 90 \% lower than trend; while for SSP5 growth is sustained at even higher pace, and stabilize at much higher levels of around 4300 passenger-kilometers per capita. For comparison, crossing AeroSCOPE \cite{salgas_aeroscope} and World Bank \cite{world_bank} averaged data for 2019: India is 198, China 902, Brazil 959, World 1377, Japan 2669, France 3561, Germany 3718, and USA 6974.

As highlighted in the assessment by Frank \textit{et al.} \cite{franz_wide_2022} when estimating demand there is also the need of endogenizing SSP narratives beyond accounting for scenario-specific GDP and population growth, a proposition is also made for incorporating \textbf{voluntary demand restraint} (for SSP1 scenarios) \textbf{or incentives} (in SSP5 scenarios) as well as an extra policy measure of market-specific demand caps for dealing with strict carbon budget constraints. Comparing these results with the ones generated by the present work, SSP1 shows a very close agreement both in terms of per-capita traffic as for total demand, SSP2 shows close agreement up until 2070, after that the logistic model employed yields lower per-capita traffic than Franz's model, with declining income elasticities. This disagreement is further intensified in SSP5, even with the higher stabilization traffic levels.

\subsubsection{Aircraft Design}

\begin{figure*}
    \centering
    \includegraphics[width=0.8\textwidth]{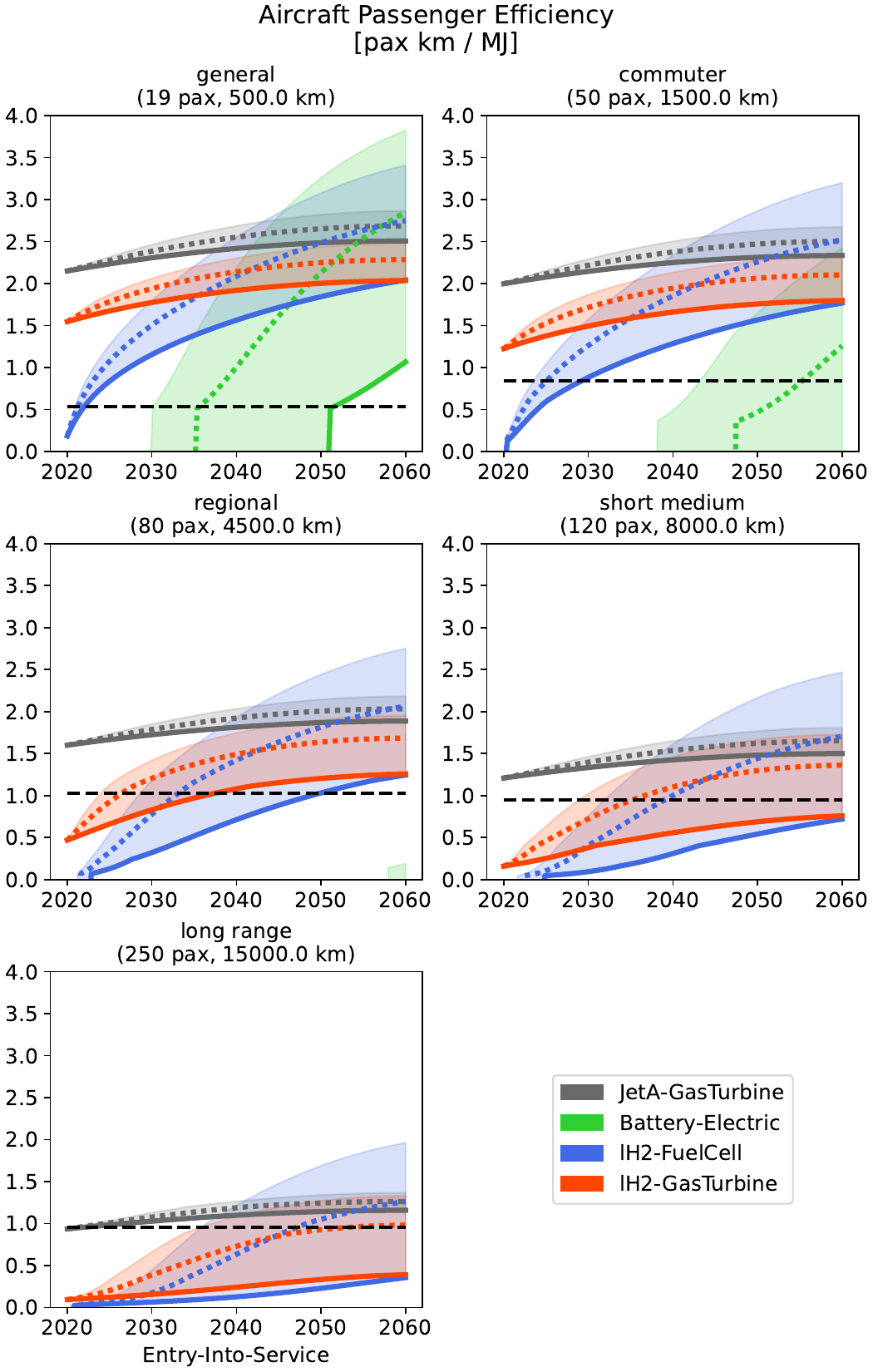}
    \caption{Expected passenger efficiency (inverse energy consumption) of prospective aircraft with technology available by Entry-Into-Service. Color-code is used to differentiate among aircraft architectures, filled between the upper and lower limit for the technology scenarios, solid line shows lower technology scenario, dotted line shows mid technology scenario. The black dashed line shows the performance of the 2019 mean fleet within the considered distance-bands \cite{salgas_aeroscope}.}
    \label{fig:aircraft-performance}
\end{figure*}

Figure \ref{fig:aircraft-performance} shows the expected design performance of aircraft design architectures depending on prospective technology scenario. Energy consumption of new designs are compared with the 2019 mean consumption from commercial flight operated within the category distance bands, obtained from the AeroSCOPE dataset \cite{salgas_aeroscope}.

\textbf{Conventional aircraft}

In general, new conventional gas turbine designs, powered by Jet-A perform better than the mean 2019 fleet, and this gap is wider over smaller distances. The diverging expectations in weight reduction expectations, yield a variation in the energy consumption of these designs depending on the chosen technology scenario, but this variation is relatively small compared to that of alternative designs.

\textbf{Electric aircraft}

Due to the low specific energy of batteries compared to other energy carriers, electric aircraft are still limited in range. Performance varies greatly with technology scenario, due to variation of batteries, electric motors, and power electronics.

Low technology yields that feasible designs in the general market are only possible around 2050. With mid and upper technology, respectively, feasibility can be achieved by 2037 and 2032 in the general market, and by 2046 and 2038 in the commuter. By 2060, with upper technology, electric aircraft is the most efficient for the general market.

\textbf{Hydrogen aircraft}

Hydrogen aircraft are feasible across all markets regardless of EIS, both for architectures that burn hydrogen in gas turbines as for using them with fuel-cells and electric motors. Technology scenario affects them more than conventional aircraft, but less than electric. The sensibility regarding technology scenario is due to the gravimetric efficiency of liquid fuel tanks, but the fuel cell architecture displays higher sensitivity, also due to electric motors and fuel cells (both in terms of efficiency and specific power).

With lower technology, burning hydrogen in gas turbines is more efficient than using fuel cell for all markets, regardless of  EIS. In the general and commuter market, as early as 2030 both are already more efficient than the reference 2019 aircraft. In the regional market, this shift happens by early 2034 and 2042, respectively. In the short-medium, by 2044 and 2050. But neither of them are able to reach 2019 efficiencies in the long-range even by 2050.

With mid technology, fuel cells surpass combustion around 2040, a bit earlier for shorter distances and a bit later for longer distances. Within the time-frame, both are able to surpass 2019 efficiencies, but neither are able to surpass the efficiency of turbofan architectures.

Finally, with upper technology hydrogen combustion is able to reach the lower-end of efficiency of conventional designs, but not the upper. Also with upper technology scenario, fuel cells are the most efficient option among architectures, except for the general market, but this supposes switching membranes with low temperature operation to high temperature \cite{ati_fuelcell}.

\subsubsection{Policy optimization}

Optimal mitigation policies are formulated as the general nonlinear programming problem. Two formulations of optimal mitigation scenarios are proposed: one relying on minimizing cumulative emissions by \textbf{supplying for trend demand}, while changing energy carriers and aircraft within the global fleet, here called trend formulation; and other that \textbf{partially avoids trend traffic} and minimizes the relative ticket price increase in order to align to a target cumulative emissions, here called low-demand formulation.



The optimization variables consist of aircraft market penetration parameters (per aircraft and per market), energy mix parameters (shares of energy production per pathway per period), and, for the low-demand formulation, the supply cap parameters (share of avoided demand per market per period). Entry-Into-Service is bounded between 2035 and 2060 for all aircraft types, except for the first generation of conventional aircraft (which can be launched and deployed as soon as 2030), the energy mix variables between 0 and 1, and $SR$ between 0 and 0.9 (at least 10 \% of the trend demand is satisfied at each market).

The optimization constraints applied to all scenarios consist in limiting the sum of the control inputs among aircraft types for each market to 1, limiting the share of pathway production to positive values among each produced energy, and limiting the consumption of biomass and electricity to a fixed fair share of the global production. Also, in the case where scenarios add electric aircraft to the fleet, an extra constraint on aircraft energy consumption is added to ensure designs are feasible by EIS.

In the trend formulation the objective function is the cumulative CO2 emissions. In the low-demand formulation, cumulative emissions are constrained (Eq. \ref{eq:co2_constraint}), and the objective is the time-discounted burden of demand aversion (Eq. \ref{eq:burden-avoidance}).

Numerical optimization is performed with the SLSQP Algorithm \cite{kraft1988software}, using forward-mode AD to obtain the gradients of objective and constraints. Forward-mode is chosen over reverse-mode, or backpropagation, due to the larger memory footprint when multiple constrains are involved (which is the case here) \cite{blo}.

\subsubsection{Robustness to background scenario}

The robustness to the background scenario was also performed as a policy optimization problem. First a formulation must be chosen (results demonstrate the case for the trend formulation) and is performed in a multi-scenario simulation. The objective of the optimization problem is then considered as the mean of policy objectives under the scenario ensemble.

The optimization variables are then divided into two sets: one that is kept fixed among all scenarios (the aircraft mix variables), and one that may be specific to each scenario (the energy mix variables). This allows to ensure the robustness of the choice of aircraft mix while allowing for each scenario to individually optimize the energy production system in order to respect their respective availability constraints.




\begin{table*}[width=\linewidth,cols=7,pos=h]
    \caption{Benchmark over a Drop-in policy optimization with trend formulation. Performed using a single Intel(R) Core(TM) i7-10850H CPU core with 2.70 GHz, as the mean of 3 repetitions, varying technology assumptions. The objective (cumulative CO2 emissions) is normalized by the 3\% of the 2°C carbon budget.}
    \centering
    \begin{tabular}{c|c|c|c c c|c}
    \toprule
     Method & Algorithm & Gradient-based & Iterations & Total time (s) & Speedup & Objective value \\
    \midrule
    
    Standard (FD) & \multirow{3}{*}{SLSQP} & \multirow{3}{*}{Yes} & X & X & X & X \\
    FD + JIT & &  & 91 & 69.3 & 87 & 1.72 \\
    JAX version &  &  & 86 & 53.2 & 114 & 1.72 \\
    \midrule
    Standard & \multirow{2}{*}{COBYLA} & \multirow{2}{*}{No} & 1852 & 6052 & - & 1.75 \\
    JAX version &  &  & 1898 & 51 & 119 & 1.75 \\
    \bottomrule
    \end{tabular}
    \label{tab:optimization}
\end{table*}

\subsection{Numerical Methodology}



The optimization of mitigation scenarios requires libraries that allow for: assembly and integration of numerous heterogeneous models; scaling with high-dimensional variables, due primarily to the disagregation among periods, technologies, and markets; minimal extra implementation when incorporating or modifying models.

Multidisciplinary Optimization (MDO) frameworks are classically used to optimize or improve a given design under various constraints \cite{mdobook}, it offers practical methods to handle model integration and coupling. Efficient optimization under high-dimension can also be achieved by using gradient-based algorithms, but their use often implies an extra burden due to the manual implementation of the derivatives of objective and constraints with regards to optimization variables. GEMSEO is an open-source Python software to automate multidisciplinary processes \cite{gallard_gems_2018}, which provided the following features that were used in the present study: automatic handling of coupled derivatives, automatic assembly of complex MDO processes based on dependency graphs, interfaces to optimization algorithms, results visualization, data storage.

Differential Programming, on the other hand, is commonly used for machine learning and scientific computing research. This programming paradigm allows for the Automatic Differentiation (AD) of lines of code. The use of AD for MDO can significantly reduce implementation burden associated to efficient high-dimension optimization. JAX is a library for array-oriented numerical computation, with capabilities for AD and just-in-time (JIT) compilation \cite{jax2018github}, enabling high-performance scientific computing in multiple hardware configurations (CPU, GPU, and TPU).

GEMSEO-JAX \cite{gemseo_jax} is an open-source plug-in that was developped by the authors to bridge JAX programs into a GEMSEO process.

\subsection{Computational gains}
\label{subsec:gains}

\begin{table}[width=\columnwidth,cols=3,pos=h]
    \caption{Benchmark over a single scenario computation and linearization. Performed using a mean of 50 repetitions in a single Intel(R) Core(TM) i7-10850H CPU core with 2.70 GHz.}
    \centering
    \begin{tabular}{c|c c}
    \toprule
    Method & Computation (s)  & Linearization (s)\\
    \midrule
    Standard (FD) & 372.7 & 362.6 \\
    JAX version & 0.45 & 0.48 \\
    \midrule
    Speedup & 830 & 749 \\
    \bottomrule
    \end{tabular}
    \label{tab:execution-linearization}
\end{table}

To evaluate the simulation gains offered by a Differential Programming paradigm, we first compare execution times over full scenario optimizations, then over a single computation and linearization.

In Table \ref{tab:optimization}, entire scenario optimization is performed with varying optimization algorithm, differentiation strategies, and compilation. Gradient-based algorithms not only require far fewer iterations for convergence but also achieve higher objective precision, confirming their advantages for high-dimensional problems \cite{perez_pyopt_2012}. Using an uncompiled model with standard Finite Differences (FD) was prohibitive due to memory demands; JIT compilation makes FD feasible yet still less efficient—both in iteration count and runtime—than forward-mode AD. Overall, using JAX yields speedups of up to two orders of magnitude for both algorithms tested.

\begin{table*}[htbp]
\centering
\setlength{\tabcolsep}{2pt}
\rotatebox{90}{
    \begin{minipage}{\textheight}
    \caption{Overview of simulated scenarios and their assumptions regarding: the background scenario, personal demand saturation, presence of traffic aversion policies, optimization objective, energy carriers included, the share of global energy production allocated to aviation, and the technology scenarios. In case multiple background scenarios, the objective is the mean among realizations. The demand saturation indicates storyline-specific assumptions on the stabilization level of per-capita demand. In presence of extra price-based traffic aversion, the objective is the minimization of relative ticket price increase. The energy carriers included are subject to variable Entry-Into-Service, and are limited to a fixed share of global production of biomass and electricity. Finally, the technology scenarios are determinant of the aircraft technology parameters, impacting the performance of new aircraft designs, driving the choice of which architectures to deploy.}
    \footnotesize
    \begin{tabular}{c|c|c c|c|c c|c}
    \toprule
    \multirow{2}{*}{Scenario name} & \multirow{2}{*}{Background scenario} & \multicolumn{2}{c|}{Demand} & Policy objective & \multicolumn{2}{c|}{Energy} & \multirow{2}{*}{Tech scenarios} \\
    & & Saturation & Traffic aversion & ($\min$) & Carriers & \% of production &\\
    \midrule
    Baseline SSP1 & SSP1-1.9 & Trend -10\% & - & Cumulative CO2 &  Jet-A (fossil) & - & Lower, mid, upper\\
    Baseline SSP2 & SSP2-2.6 & Trend & - & Cumulative CO2 &  Jet-A (fossil) & - & Lower, mid, upper\\
    Baseline SSP5 & SSP5-4.5 & Trend +50\% & - & Cumulative CO2 & Jet-A (fossil) & - & Lower, mid, upper\\
    \midrule
    Drop-in trend & SSP2-2.6 & Trend & - & Cumulative CO2 &  Jet-A (fossil+SAF) & 5.0 & Lower, mid, upper\\
    Drop-in availability & SSP2-2.6 & Trend & - & Cumulative CO2 &  Jet-A (fossil+SAF) & 8.6 & Lower, mid, upper\\
    Drop-in low-demand & SSP2-2.6 & Trend & \checkmark & \textbf{Rel. price increase} &  Jet-A (fossil+SAF) & 5.0 & Lower, mid, upper\\
    \midrule
    Breakthrough trend & SSP2-2.6 & Trend & - & Cumulative CO2 & Jet-A (fossil+SAF), LH2, Battery & 5.0 & Lower, mid, upper\\
    Breakthrough availability & SSP2-2.6 & Trend & - & Cumulative CO2 & Jet-A (fossil+SAF), LH2, Battery & 8.6 & Lower, mid, upper\\
    Breakthrough low-demand & SSP2-2.6 & Trend & \checkmark & \textbf{Rel. price increase} & Jet-A (fossil+SAF), LH2, Battery & 5.0 & Lower, mid, upper\\
    \midrule
    Scenario-robust trend & SSP2-1.9, 2.6, and 3.4 & Trend & - & \textbf{Mean} cumulative CO2 & Jet-A (fossil+SAF), LH2, Battery & 5.0 & Mid\\
    \bottomrule
    \end{tabular}
    \label{tab:scenarios}
    \end{minipage}
}
\end{table*}

In Table \ref{tab:execution-linearization} the computation compares compares how long one single scenario computation takes to complete with (JAX version) and without (standard) JIT compilation, and the time required to linearize objectives and constraints. JAX introduces an initial overhead of roughly 38\,s due to compilation; this is excluded from single computation comparisons but included in full-optimization timings. Once compiled, JAX achieves speedups of up to three orders of magnitude.

For context, a single scenario simulation using a restricted subset of AeroMAPS models \cite{planes_aeromaps_2023} (default bottom-up setup without offsetting, non-CO\textsubscript{2}, or cost models) takes 1.29\,s, about three times slower than the present approach. This comparison should be interpreted cautiously, as the modeling scopes differ substantially: AeroMAPS includes freight, non-CO\textsubscript{2} effects, and cost models, but omits aircraft design routines and links between traffic growth and socioeconomic drivers.

\section{Scenario Results}
\label{sec:results}

Several policy scenarios are simulated using the numerical optimization methods presented, these are summarized in Table \ref{tab:scenarios}.

We start with the \textbf{baseline} no-policy scenarios (SSP1, 2, and 5), where two new generations of conventional aircraft are launched, and their entry-into-service and deployment is optimized to minimize cumulative emissions, while consuming only fossil kerosene. The sensibility to maturing aircraft technology is also explored with 3 technology scenarios with explicit time-varying component-level performances (see Figures \ref{fig:aircraft-performance} and \ref{fig:aircraft-tech}), impacting energy consumption of new aircraft.

\begin{figure*}[htbp]
    \centering  
    \includegraphics[width=\textwidth]{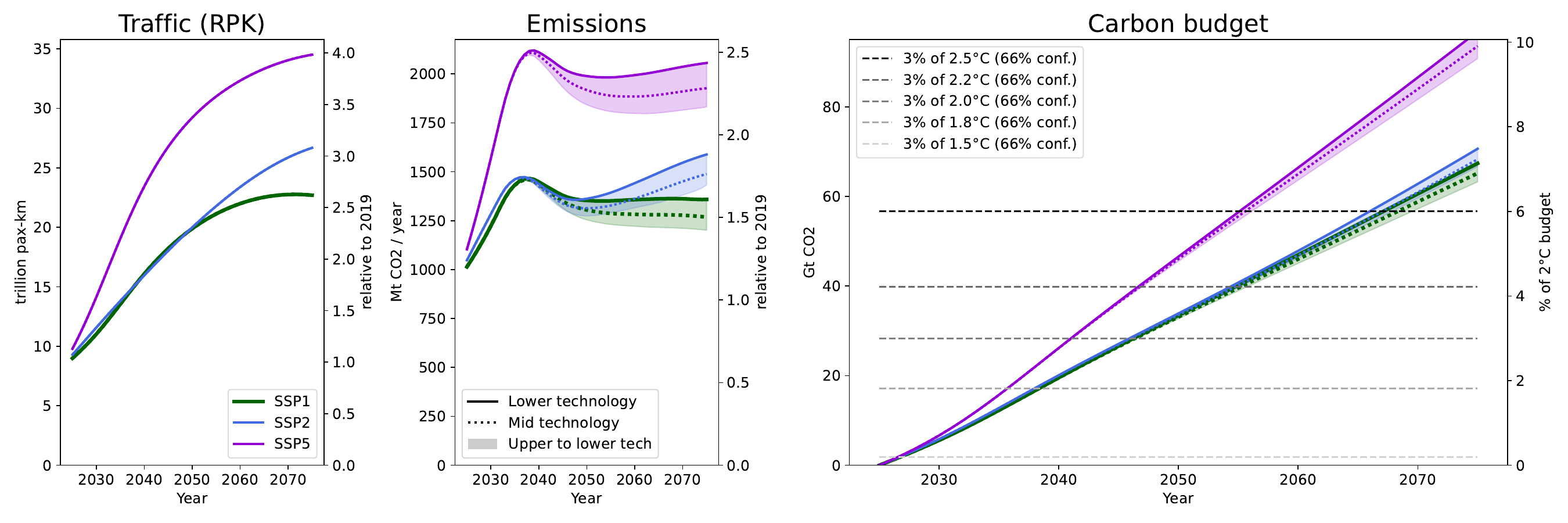}
    \caption{Comparison of traffic, annual emissions, and cumulative emissions of no-policy baseline scenarios SSP1 (dark green), SSP2 (blue), and SSP5 (purple). The sensibility to aircraft technology is displayed with 3 aircraft technology scenarios: Lower technology (continuous line), Mid technology (dotted line), and Upper technology (shadowed region).}
    \label{fig:baseline}
\end{figure*}

Then the mitigation scenarios are explored using SSP2 as the baseline. The Drop-in \textbf{trend} mitigation scenarios also introduce incorporation of biofuel and electrofuel (SAF) in the Jet-A blend, and constraint the sectoral consumption of electricity and biomass to 5.0 \% of the global supply. The Drop-in \textbf{availability} mitigation scenario increases the sectoral consumption to 8.6 \%. And the Drop-in \textbf{low-demand} mitigation scenario keeps trend consumption, but avoids traffic in order to fulfill an additional constraint on the total cumulative emissions. As the baseline scenarios, these are also explored with 3 technology scenarios.

The Breakthrough mitigation scenario increments the Drop-in by introducing new alternative aircraft concepts (Battery-Electric, LH2 Fuel-Cell, and LH2 Gas Turbine), and is also divided into a trend, availability, and low-demand variant, each sweeping the 3 technology scenarios.

Finally, the scenario-robust mitigation scenarios keep all mitigation measures (SAF and deployment of alternative aircraft), but the objective is now to optimize the mean among 3 different background scenarios: SSP2-1.9, 2.6, and 3.4. These scenarios keep the trend assumption of 5.0 \% global energy production allocated to aviation, and are divided into a trend and low-demand variant. For simplification purposes, these are only explored with the mid aircraft technology.

\subsection{Baseline scenarios}

The emissions in baseline scenarios are mainly driven by the growth in RPK demand, shown in Figure \ref{fig:baseline} along with the annual and cumulative CO2 emissions. SSP5 displays the highest initial demand growth, while SSP1 and 2 show close agreement until around 2050, where SSP2 demand continues to grow while SSP1 demand peaks and then declines.

All scenarios are subject to the deployment of new conventional aircraft by 2030, which allows for reducing associated carbon-intensity of demand. This leads to annual emissions that are growing until a peak is reached between 2035 and 2040, where efficiency gains due to aircraft replacement outpace demand growth. However, after the introduction of the first generation of new aircraft, further energy consumption gains with the second generation are only marginal. In SSPs 1 and 5 emissions practically stabilize after the peak due to due to slower pace of demand growth, while continued demand growth in SSP2 result growing emissions after fleet is replaced. All of the baseline scenarios display much higher emission levels compared to the year 2019, around 50 \% higher for SSP1, 80 \% for SSP2, and 120 \% for SSP5.

\begin{figure*}[htbp]
    \centering
    (a) Drop-in mitigation:
    
    \includegraphics[width=\textwidth]{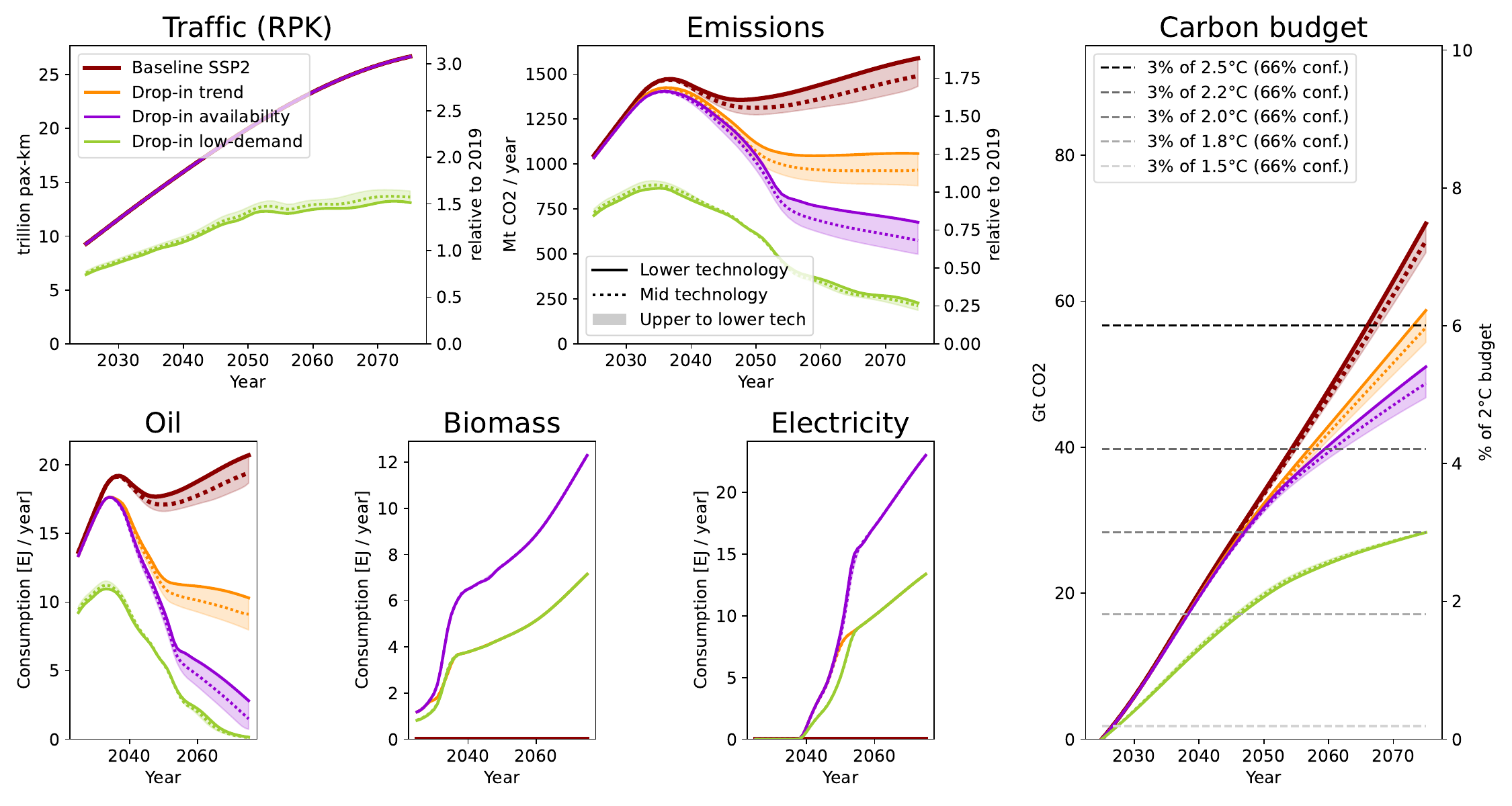}

    \bigskip

    (b) Breakthrough mitigation:

    \includegraphics[width=\textwidth]{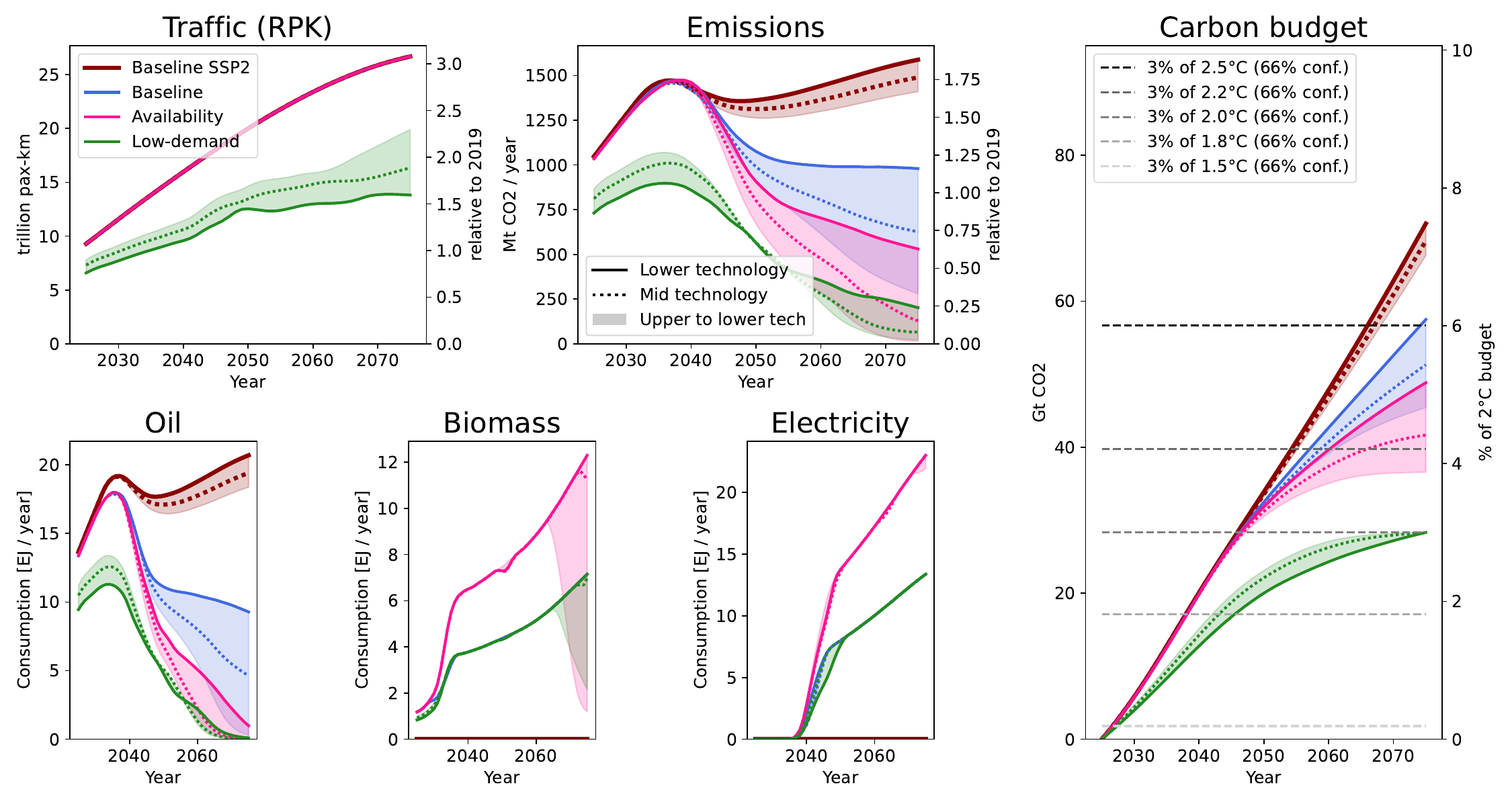}
    
    \caption{Comparison of drop-in (a) and breakthrough (b) mitigation scenarios under: trend, trend with extra energy availability, and low-demand formulations. The sensibility to aircraft technology is displayed with 3 scenarios of component-level performances: Lower technology (continuous line), Mid technology (dotted line), and Upper technology (shadowed region). These also impact the performance of new aircraft designs, driving the choice of which architectures to deploy and when they are launched.}
    \label{fig:mitigation}
\end{figure*}

Cumulative emissions are also much higher than what the considered fair share for aviation. All baseline scenarios exceed its share of contribution to the Paris Agreement target of 2° C by a factor of 2-3, exceeding its share of carbon budget as soon as 2040 in SSP5 and 2046 in SSPs 1 and 2.

\subsection{Mitigation policy optimization}

The outcomes of Drop-in and Breakthrough mitigation are presented in Figure \ref{fig:mitigation} and compared with the no-policy baseline SSP2, named hereafter as "Fossil" scenario.

\begin{figure*}
    \centering
    (a) Drop-in trend, mid tech:
    
    \includegraphics[clip, trim=0cm 9.8cm 0cm 0cm, width=0.85\textwidth]{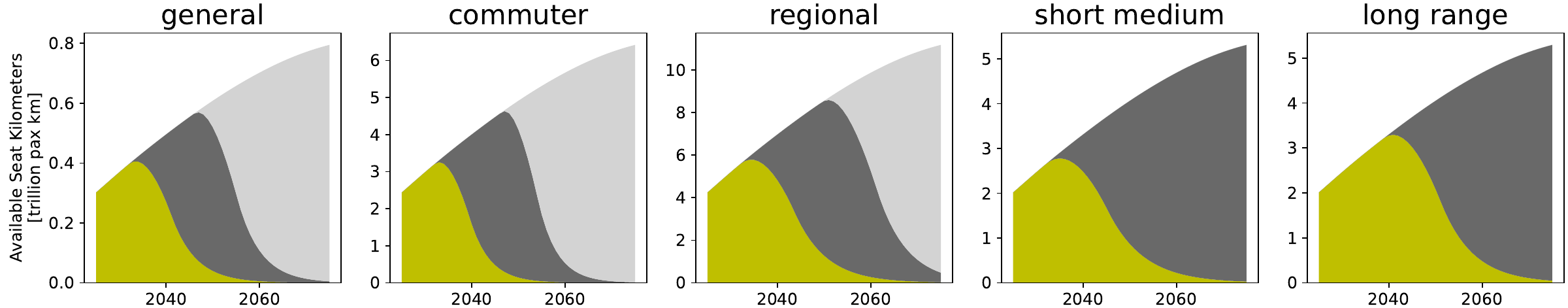}
    \bigskip

    (b) Drop-in low-demand, mid tech:

    \includegraphics[clip, trim=0cm 10cm 0cm 0cm, width=0.85\textwidth]{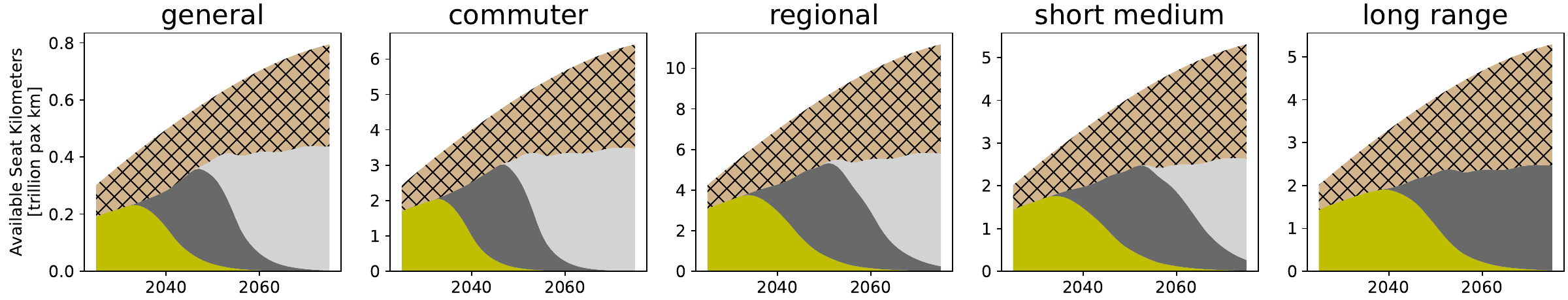}
    \bigskip

    (c) Breakthrough trend, lower tech:

    \includegraphics[clip, trim=0cm 10.6cm 0cm 0cm, width=0.85\textwidth]{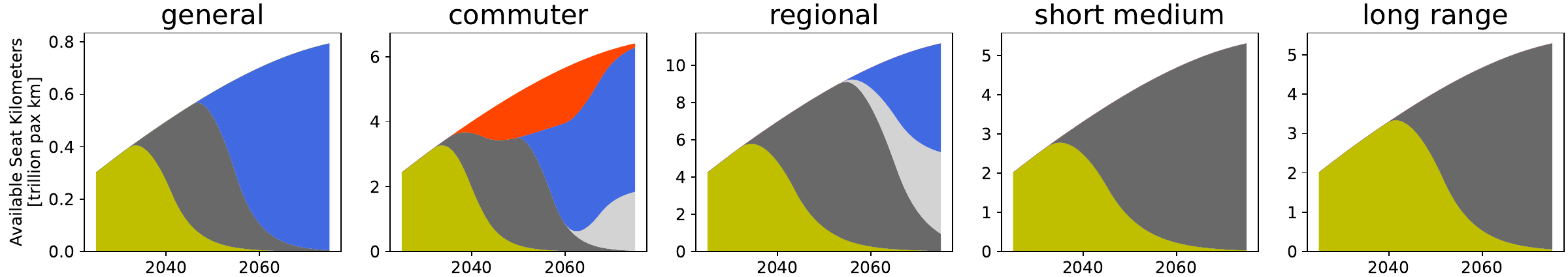}
    \bigskip

    (d) Breakthrough availability, lower tech:

    \includegraphics[clip, trim=0cm 10.6cm 0cm 0cm, width=0.85\textwidth]{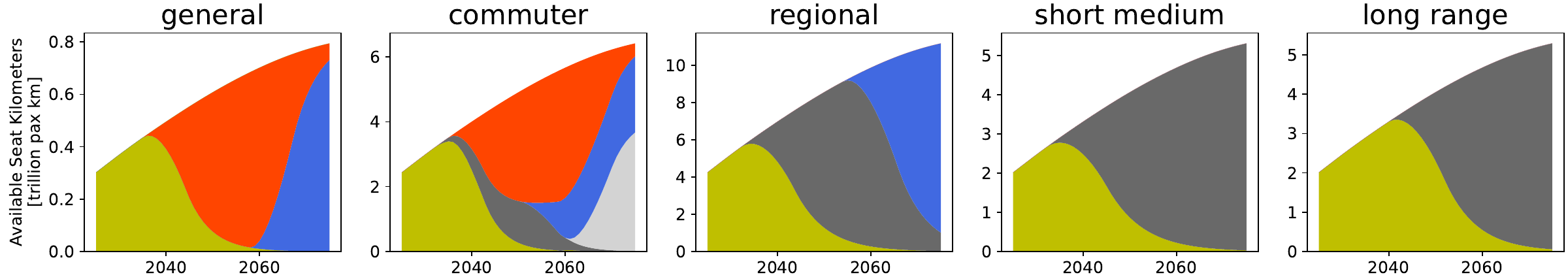}
    \bigskip

    (e) Breakthrough trend, upper tech:

    \includegraphics[clip, trim=0cm 10.6cm 0cm 0cm, width=0.85\textwidth]{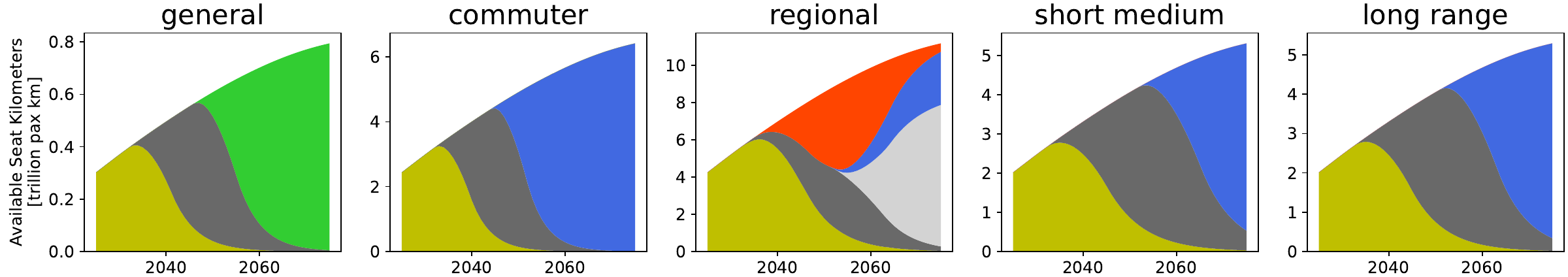}
    \bigskip

    (f) Breakthrough low-demand, mid tech:

    \includegraphics[clip, trim=0cm 6.3cm 0cm 0cm, width=0.85\textwidth]{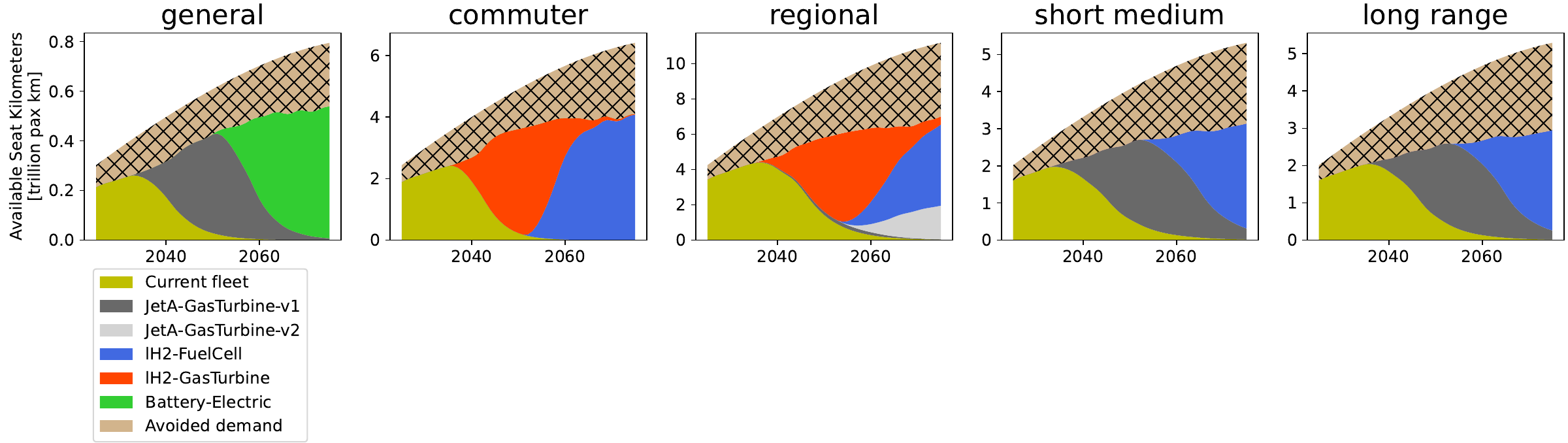}
    \bigskip
    
    \caption{Comparison of aircraft fleet composition per market for a set of simulated mitigation scenarios. Assumptions on energy availability and aircraft technology significantly impacts the performance of new aircraft designs, driving the choice of which architectures to deploy and when they are launched.}
    \label{fig:fleet}
\end{figure*}

Overall, the Drop-in mitigation scenarios are less sensitive to aircraft technology assumptions, as emission reductions is achieved either with SAF incorporation or demand aversion. The trend variant reaches emission levels comparable to 2019, but fails to phase out of consuming fossil fuels. The availability variant further reduces emission levels, and allow for much lower fossil consumption. The low-demand variant, traffic is nearly half of the trend by 2070, but still around 50 \% higher than the 2019 reference, this variant is the only one that allows for keeping the cumulative emissions in check with the Paris Agreement, phasing out of fossil consumption, and lowering the emissions peak.

The Breakthrough mitigation displays much higher sensitivity to aircraft technology, lower emission levels than Drop-in mitigation (when compared with similar technology), and lower fossil consumption. The trend variant, for instance, still consumes fossil fuels and can emit more than 2019 emissions with lower technology, or less than half of 2019 with upper technology. The availability variant can reach near zero emissions after 2070, phasing out of fossil can happen as soon as 2060, allowing for relaxing the biomass consumption constraint, but cumulative emissions are still higher than the sector's fair share. The low-demand variant respects the carbon budget constraint with much less difficulty than the Drop-in low-demand, allowing for much higher demand levels, which may reach up to 140 \% more traffic than 2019, but is still lower than the trend 200 \% increase.

Regarding biofuel production, while the HEFA pathway displays significantly higher emissions compared to FT, it consumes significantly less biomass. This yields that, in scenarios where fossil kerosene consumption is still high, biomass is preferentially allocated to HEFA production. For electrofuel production, Breaktrough scenarios display much lower shares of electrofuel in the Jet-A blend, because hydrogen and electricity are preferentially allocated to alternative aircraft rather than to make electrofuel, but this trade-off is highly dependent on the flight distance considered.

\begin{figure*}[htbp]
    \centering
    \includegraphics[width=\textwidth]{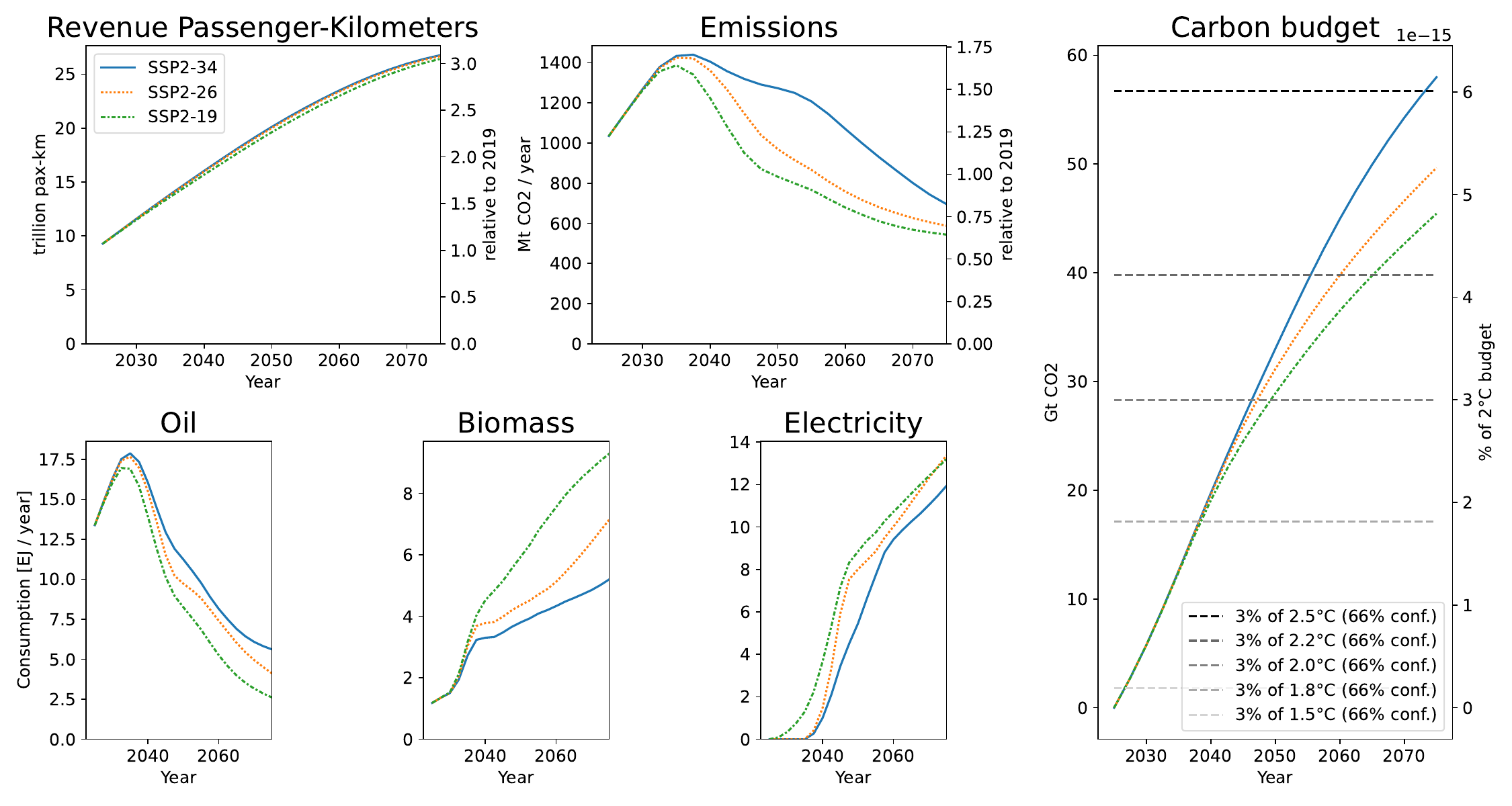}
    \caption{Comparison of scenario-robust mitigation under Mid aircraft technology and background scenarios SSP2 with RCP's 1.9, 2.6, and 3.4. These strategies prioritizes flexibility by sizing alternative-aircraft deployment for the worst electricity conditions (SSP2-3.4), yielding higher SSP2-2.6 emissions but greater adaptability across SSP2-1.9 and SSP2-3.4.}
    \label{fig:robust}
\end{figure*}

Regarding the aircraft fleet, Figure \ref{fig:fleet}) provides a view of how aircraft technologies are composed to make up the supply in each market segment in some of the simulated scenarios. The general and commuter markets are consistently the first to transition toward new aircraft, whereas short-medium and long-range markets remain dependent on drop-in fuels for longer periods. The Breakthrough scenarios (c–e) demonstrate the importance of both technological maturity and availability assumptions: lower technology (c, d) delay deployment of hydrogen systems, but once extra energy availability is assumed (d), alternative aircraft are launched as soon as available. In contrast, the upper technology case (e) allows for more aggressive displacement of conventional aircraft, and also drive the adoption of Battery-Electric in the general market and of Hydrogen Fuel-Cells on remaining markets. The low-demand scenarios (b, f) reduce the scale of alternative propulsion adoption allowing for hydrogen aircraft in the short-medium and long-range markets even with trend energy availability.

\subsection{Scenario-robust mitigation policy}

\begin{figure*}
    \centering

    (a) SSP2-1.9:
    
    \includegraphics[width=0.7\textwidth]{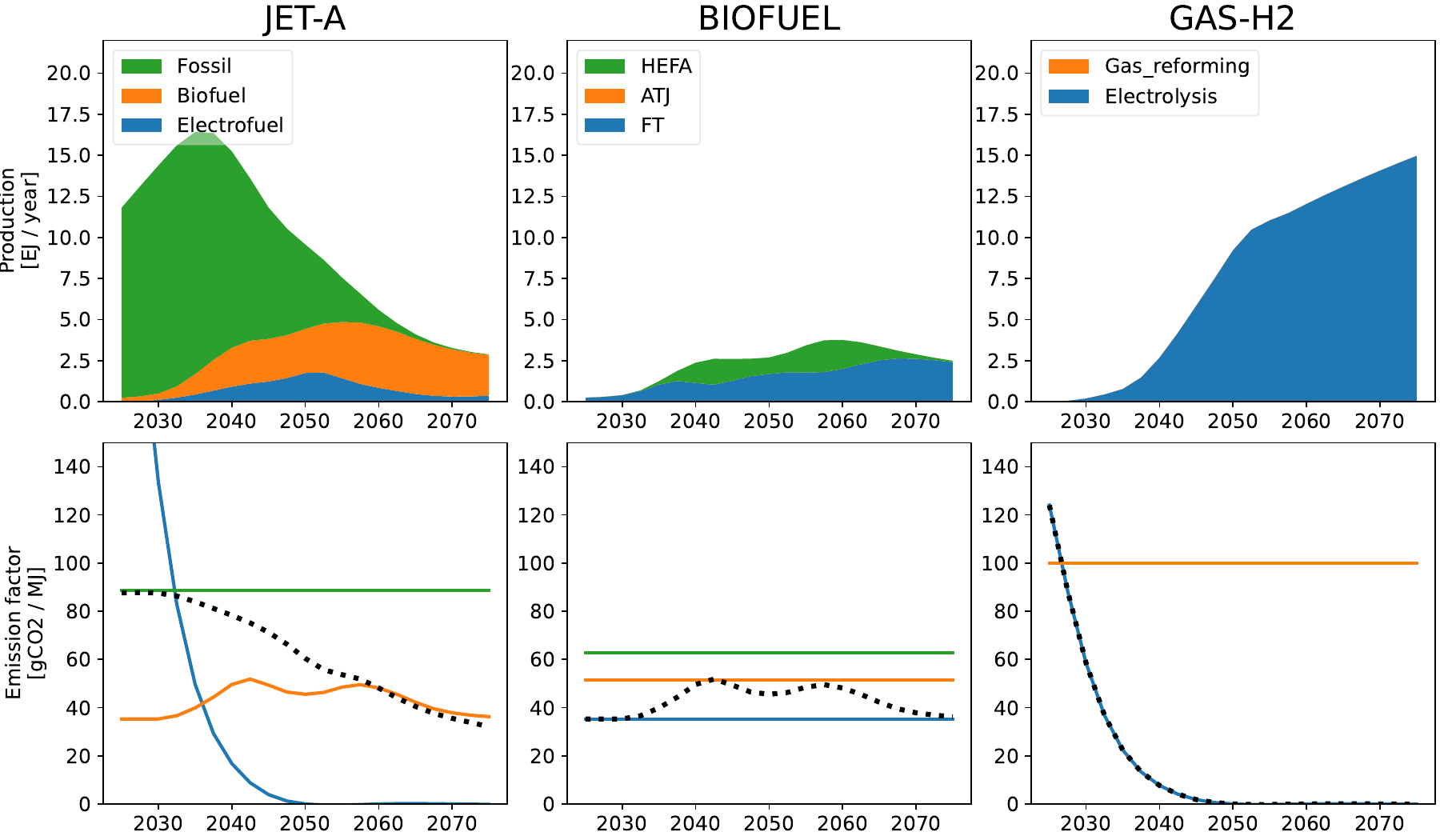}

    (b) SSP2-2.6:
    
    \includegraphics[clip, trim=0cm 8.5cm 0cm 0cm, width=0.7\textwidth]{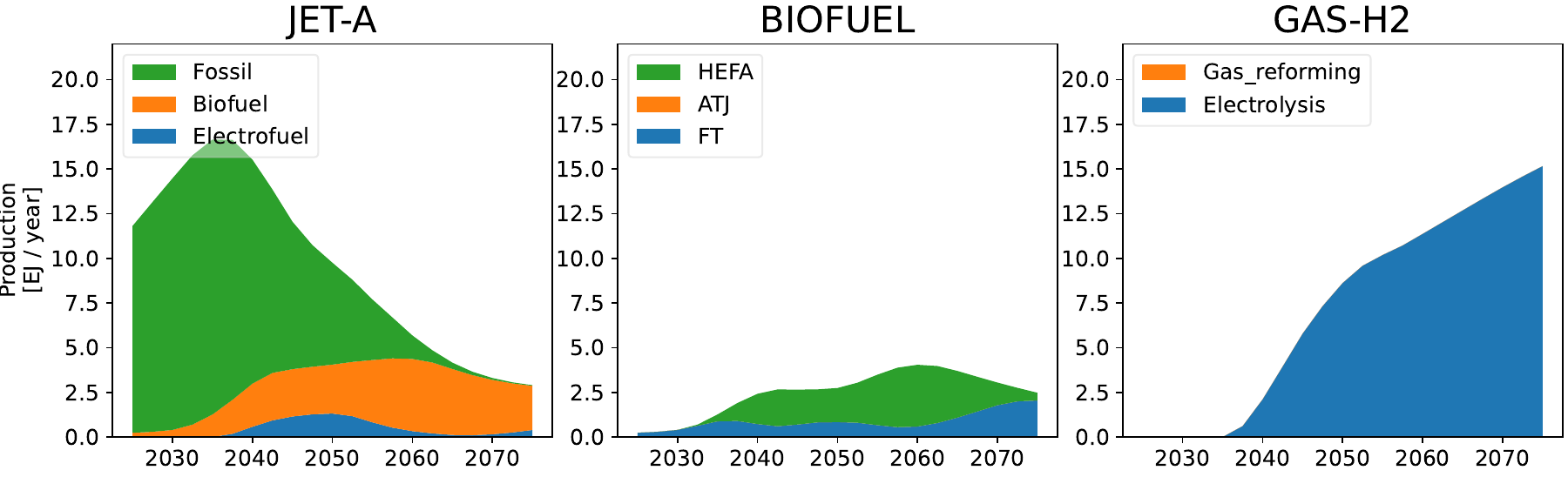}

    (c) SSP2-3.4:
    
    \includegraphics[width=0.7\textwidth]{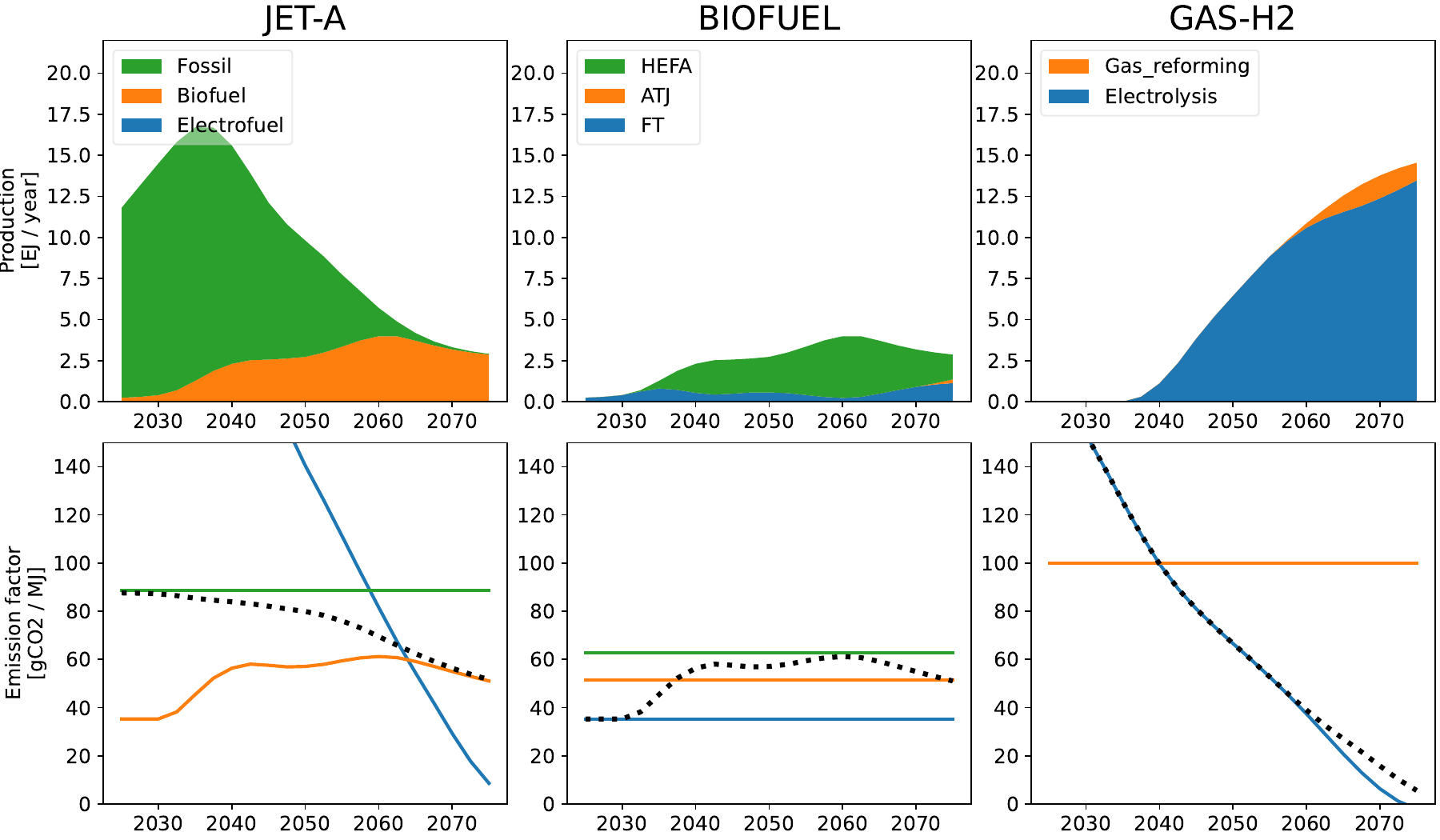}
    
    \caption{Energy mix under scenario-robust trend mitigation for multiple background scenarios: SSP2 with 1.9, 2.6, and 3.4 forcing levels. The top graph shows how energy production is split among different pathways. The bottom graph shows the emission factor of each pathway (continuous line), and the mean emission factor after mixing production from all pathways (dotted line).}
    \label{fig:robust-energy}
\end{figure*}

As the SSP2-2.6 was chosen as background scenario for the single policy optimization, we include scenarios SSP2-1.9 and SSP2-3.4 in the scenario-robust optimization, to account for different radiative forcing targets, representing a more ambitious and a less ambitious scenario. The outcomes of this multi-scenario policy optimization are shown in Figure \ref{fig:robust}. Compared to the single-scenario optimum, the robust optimum display higher emissions on the target scenario SSP2-2.6, yet it allows for greater flexibility in the case an adjacent global scenario takes place. Overall, this is due to smaller shares of alternative aircraft, driven by the strong difference in the availability of electricity and emission factor associated to grid electricity among the scenarios. The overall strategy is to use the worst case scenario (higher electricity emission factor and less electricity availability) to determine the deployment of alternative aircraft, and use the surplus electricity in the other scenarios to make extra electrofuels for conventional aircraft, leaving room for more variability.

Figure \ref{fig:robust-energy} shows the energy mix in the SSP2 scenarios with 1.9, 2.6, and 3.4 forcing levels. Among these three scenarios, lower warming levels lead to: earlier deployment of electrofuels due to lower electricity emission factor, higher shares of electrofuel due to the higher electricity availability, higher shares of FT pathway in the biofuel blend, and higher shares of biofuel in the Jet-A blend.

\subsection{Comparison with literature}

\begin{figure*}
    \centering
    \includegraphics[width=\textwidth]{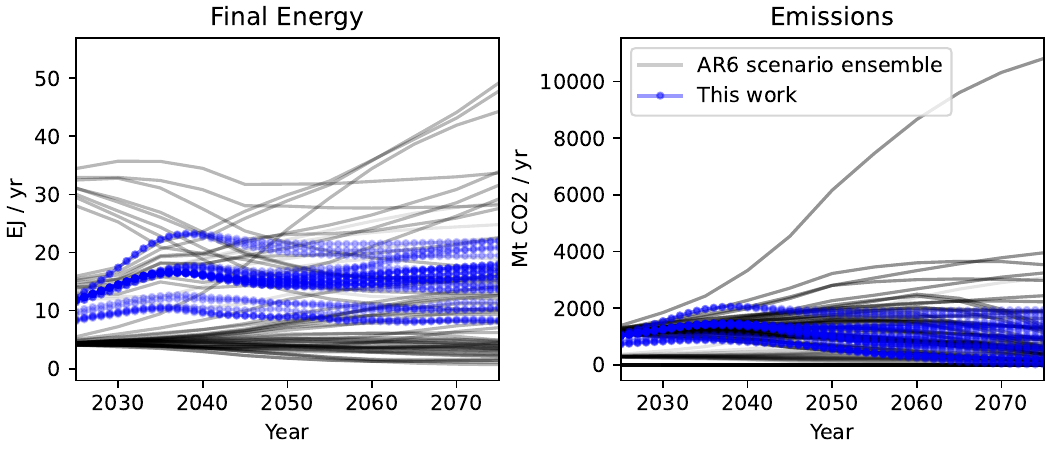}
    \caption{Comparison of aviation final energy demand and direct CO2 emissions between the scenarios developed in this work and the AR6 scenario ensemble \cite{ar6_database}. The AR6 shaded envelope reflects diverse assumptions across models regarding aviation scope, demand growth, efficiency gains, and fuel transitions. The trajectories from this study remain within the ensemble’s range through mid-century, while the stronger long-term decline results from explicitly modeling per-capita demand saturation and full fleet renewal.}
    \label{fig:comparison}
\end{figure*}

Figure \ref{fig:comparison} contrasts aviation's projected final energy use and emissions from the scenarios generated in this work against the AR6 scenario ensemble. The AR6 ensemble exhibits substantial variation, reflecting divergent modeling assumptions regarding aviation scope (international versus domestic, commercial operations, freight inclusion), demand growth trajectories, efficiency improvements, and alternative fuel adoption pathways. The scenarios developed in this study fall within the inter-model spread of the AR6 ensemble for both energy consumption and CO2 emissions through approximately 2050, demonstrating broadly consistent near- to mid-term dynamics.

Beyond 2050, the present work projects lower energy and emissions compared to much of the AR6 ensemble. This divergence stems primarily from the incorporation of demand saturation effects—as per-capita incomes rise according to SSP narratives, the logistic demand model captures the stabilization of air travel propensity at high income levels, rather than assuming continued growth. This comparison validates the plausibility of the scenarios presented in this work, as they remain consistent with the AR6 range while providing additional insight through explicit modeling of demand saturation dynamics, technology-specific aircraft performance, and resource constraints.

\section{Limitations}
\label{sec:limits}

This section describes the limitations of the numerical methods and the simplifying hypothesis made by the models used in the present paper.

Firstly, using differential programming paradigms requires implementing model equations in specific frameworks, which may be stricter than original implementations that don't allow for AD. While the results themselves are not affected, the limitation here is related to rethinking parts of implementation, especially for conditional statements, iterative processes, and in-place replacement.

Secondly, the limitations of the models are noted for each elementary model:

\subsection*{Air traffic demand}

The use of logistic equations, when compared to constant elasticity models, allows for better account of the economic development of countries, but using these in a forward-looking manner means exploiting the model outside of its calibrated range, which can be particularly problematic when estimating saturation levels in countries that have not yet reached the inflection point of the curve.

The scenarios here are aggregated on a global level, which does not account for regional disparities in demand growth, carbon content of electricity, and biomass availability.

Also, even though mitigation costs are not modeled in this work, prices are also a driver for demand modifications through price elasticity. Implementing this effect leads to a coupled system (energy cost depends on traffic volumes, which depends on ticket prices), which is highly dependent on policy, e.g., taxes and incentives.

\subsection*{Aircraft design}

TLAR are kept the same for each architecture within a market, but aircraft size, speed, and altitude can have affect vehicle performance and ideally should be optimized for each of them.

Cryogenic tank gravimetric efficiency matures with time, but is kept fixed among markets. This ignores dependency on tank size \cite{adler_hydrogen-powered_2023, ati_cryogenic}, the upper technology scenario may figure values that are not realistic for small aircraft.

Finally, gas turbine performance weight and consumption was kept the same between kerosene and hydrogen. Litterature shows that energy consumption may be slightly different when burning hydrogen \cite{mourouzidis_abating_2024}, but few information is available on the weight penalty of such engines.

\subsection*{Fleet deployment}

Within each market segment, aircraft architectures are tailored for the flown distances. Yet, in practice airlines may operate aircraft over distances that are different from which the aircraft was optimized for, this leads to increased fuel consumption in reality and explains the lower historical efficiencies for the general and commuter segments.

\subsection*{Energy-mix}

Because the energy production models are formulated to be "controlled" by the optimizer using a time-delay on the shares of production (rather than production volumes itself), whenever production volumes of energy carriers varyies rapidly, the pathway production struggles to adapt its shares to keep steady production levels. For example, around 2050 in scenarios with introduction of alternative aircraft, the consumption of Jet-A decreases sharply leading to some oscillations in the total biomass consumption and in the repartition of biofuel production among the modeled pathways.

No difference is made between biomass types, which means all pathways compete for the same resource. In practice, each biofuel pathway consume specific feedstocks that may not necessarily compete with each-other and subject to very different production volumes.

Finally, the electricity consumption associated with the power-to-liquid pathway (Tab. \ref{tab:energy-prod}) is representative of a process that uses solely Direct Air Capture as the CO2 source. Using concentrated CO2 sources can significantly decrease this and while some assessments conclude that some regional sources can be sufficient to ramp-up production \cite{drunert_ptl}, the assumption of no competition for the CO2 volumes can be highly questionable, especially when accounting other transportation modes.

\section{Discussion}
\label{sec:discus}

Considering trend traffic growth and allocation of energy resources, even with optimistic technological breakthrough, aviation share of emissions will grow as the wider economy decarbonizes in a 2°C temperature increase scenario. This is in line with scenarios from the IPCC's 6th Assessment Report \cite[][Figure 3.19]{long_ar6_wg3}, in which transport emissions take the longest to reach net zero even in scenarios below 1.5°C with little overshoot.

Regarding the implications for the energy system, fulfilling aviation's needs will require significant volumes of low-carbon electricity and biomass. Limited production amounts will require dedicated policy on which sectors to prioritize the access to these resources. These findings are also in line with recent studies focused on the aviation sector \cite{becken_implications_2023, eaton_regional_2024, drunert_ptl}. Dedicated production facilities integrated near airports might be a way to reducing associated supply chain losses and costs \cite{hoelzen_h2-powered_2025}.

As the efficiencies of energy conversion processes can vary strongly depending on the primary energy source, their comparison must be made on a per unit of service basis. While some studies on several transport modes \cite{wallington_green_2024} consider the base service unit as 1 MJ of thrust (or to wheels), this work instead, considers it to be a traffic measure (passenger-kilometer). This distinction is fundamental for aviation due to the different weights of the propulsion architectures to power these vehicles.

Air transportation is highly unequal sector, both across (Fig. \ref{fig:regional-demand}) and within \cite{GOSSLING2020102194} countries. Given that most of the traffic growth in the next decades will happen in emerging economies, which have not yet reached demand saturation. Many of these, however, are still steering public policy to increase access to air travel, e.g., \cite{voa-brasil}. The public acceptance and effectiveness of demand-side measures can be questioned, and using pricing mechanisms to address this problem also raises questions of social justice \cite{Buchs02012024}.

Also, climate change is but one within the Planetary Boundaries \cite{rockstrom_safe_2009, richardson_earth_2023}. Some recent assessments that expand the scope for other Planetary Boundaries \cite{pais_current_2024} stress on the need for acknowledging other limits to avoid shifting the problem, such as biodiversity loss and eutrophication of freshwater ecosystems, especially when considering a strong uptake of biofuels, which are present in most of the industry decarbonization roadmaps \cite{becken_implications_2023}.

Finally, given a set of technology forecasts, optimization can be useful for deciding which technologies to prioritize and when, especially when they compete for similar resources. Yet, in many IAM applications that use optimization, the formulation of the optimization problem is often left unchanged or is little discussed, even if the choice of policy goal (objective) and non-negotiables (constraints) are a fundamental part of the process.

\section{Conclusion}
\label{sec:conclus}

Overall results show that, under trend demand growth: baseline scenarios display a peak in emissions between 2035 and 2040, mitigation scenarios based solely on SAF has limited emissions reductions due to energy availability constraints (2070 emissions are still 75 \% of 2019 with preferential availability), breakthrough aircraft technologies can allow for reaching near zero emissions, but their impact is delayed to after 2045 due to late Entry-Into-Service and slow fleet renewal.

The choice of which aircraft architecture and energy carrier to embark is highly dependent on vehicle performance (determined by TLAR and technology assumptions) and on the background energy system (due to the timing of electricity decarbonization and to limited availability of electricity and biomass).

In order to respect Paris Agreement targets under an effort-sharing principle, drop-in mitigation will put the system to a higher stress: either by constraining traffic, or by consuming too much biomass and electricity. New aircraft with alternative energy carriers allows to reduce such stress by making better use of the same energy resources, even if their energy consumption is higher than that of conventional planes.

When comparing mitigation policies with different objective functions, it was found that supply caps, energy availability, and the introduction of alternative aircraft designs are complementary rather than competing measures. One strategy alone can achieve reduction in emissions up to a certain level, but their combined use is capable of more efficient mitigation by: using alternative aircraft in its feasible markets (reducing needs for low-carbon electricity and biomass for a given service), and avoiding emission-intense markets (further reducing consumption of fossil kerosene).

Regarding the numerical methods, the use of GEMSEO-JAX allowed for both reducing implementation burden and execution time for the optimization of mitigation scenarios. The simulation of these optimization-based policy scenarios was prohibitive without the speedups obtained, which are of two orders of magnitude at the scenario optimization level and three orders of magnitude at the scenario computation and linearization level.

Our research offers practical insights on how to efficiently use optimization algorithms for mitigation scenarios. Applying these methods for the aviation sector, also shed light on how to optimally allocate aircraft architectures, energy resources, and demand-side measures to achieve stringent mitigation targets.

\section*{Code availability}

All the scripts and data required to reproduce the results from this work are openly available in \url{https://gitlab.com/ian.costa-alves1/noads}.

\section*{Acknowledgements}

Gratitude is extended to the Conceptual Airplane Design and Operations (CADO) team at École Nationale de l'Aviation Civile (ENAC), to the Aviation, Climate, Environment (ACE) group at ISAE-SUPAERO, and to the Institute for Sustainable Aviation (ISA) for their support, assistance, and fruitful discussions. The Generic Aircraft Design Model (GAM), provided by the CADO team, was essential for enabling modeling and analysis of alternative aircraft designs. The AeroMAPS platform, developped by ISAE-SUPAERO and ISA, was reponsible for laying the groundwork upon which this research was built. Special thanks to Pascal Roches, Nicolas Monrolin, Thomas Planès, Scott Delbecq, Antoine Salgas, Florian Simatos, Laurent Joly, and Xavier Carbonneau for their sharp insights, support, and collaboration.

Also, the authors thank the Multidisciplinary Optimization Competence Center at IRT Saint Exupéry, for their availability and support with the methodological developments that preceded this research. Special thanks to Matthias De Lozzo, and Antoine Dechaume for their aid with repository maintenance and thorough code reviews.

\section*{CRediT authorship contribution statement}

\textbf{Ian Costa-Alves}: Writing – original draft, Writing – review \& editing, Conceptualization, Data curation, Investigation, Methodology, Software, Validation, Visualization.

\textbf{Nicolas Gourdain}: Writing – original draft, Writing – review \& editing, Conceptualization, Methodology, Funding acquisition, Project administration, Supervision.

\textbf{François Gallard}: Writing – original draft, Writing – review \& editing, Conceptualization, Methodology, Software, Supervision.

\textbf{Anne Gazaix}: Writing – review \& editing, Conceptualization, Funding acquisition, Project administration, Supervision.

\textbf{Yri-Amandine Kambiri}: Writing – review \& editing, Software, Validation.

\textbf{Thierry Druot}: Writing – review \& editing, Conceptualization, Software, Supervision, Validation.

\section*{Funding sources}

This work was supported by the Occitania region, ISAE-SUPAERO, and IRT Saint Exupéry.

\begin{nomenclature}
\textit{Abreviations}
\begin{deflist}[AAAAAA] 
\defitem{AD}\defterm{Automatic Differentiation}
\defitem{ASK}\defterm{Available Seat Kilometers}
\defitem{EIS}\defterm{Entry-Into-Service}
\defitem{FD}\defterm{Finite Differences}
\defitem{GAM}\defterm{Generic Airplane Model}
\defitem{GDP}\defterm{Gross Domestic Product}
\defitem{IAM}\defterm{Integrated Assessment Model}
\defitem{JIT}\defterm{Just-In-Time}
\defitem{MDO}\defterm{Multidisciplinary Optimization}
\defitem{RCP}\defterm{Representative Concentration Pathway}
\defitem{RPK}\defterm{Revenue Passenger Kilometers}
\defitem{SAF}\defterm{Sustainable Aviation Fuel}
\defitem{SSP}\defterm{Shared Socioeconomic Pathways}
\defitem{TLAR}\defterm{Top Level Aircraft Requirements}
\end{deflist}

\bigskip

\textit{Variables}
\begin{deflist}[AAAAAA] 
\defitem{$o$}\defterm{Delay output variable}
\defitem{$i$}\defterm{Delay input variable}
\defitem{$\tau$}\defterm{Delay-time (time constant)}
\defitem{$D$}\defterm{Aggregate demand}
\defitem{$Pop$}\defterm{Population}
\defitem{$I$}\defterm{Per-capita income}
\defitem{$p$}\defterm{Ticket price}
\defitem{$\epsilon_{\upsilon}$}\defterm{Elasticity of demand to variable $\upsilon$}
\defitem{$\sigma$}\defterm{Calibration constant}
\defitem{$RPKpc_{trend}$}\defterm{Trend per-capita Revenue Passenger Kilometers}
\defitem{$L$}\defterm{Left asymptote (propensity to travel at 0 income)}
\defitem{$R$}\defterm{Right asymptote (propensity to travel at $\infty$ income)}
\defitem{$\iota$}\defterm{Income per capita at the inflection point}
\defitem{$B$}\defterm{Logistic growth rate}
\defitem{$C$}\defterm{Logistic control of the transition between asymptotes}
\defitem{$\nu$}\defterm{Logistic control of maximum growth}
\defitem{$RPKpc_{story}$}\defterm{Storyline-adjusted per-capita RPK}
\defitem{$F_{SSP}$}\defterm{Storyline multiplier factor}
\defitem{$RPK_{trend}$}\defterm{Trend total Revenue Passenger Kilometers}
\defitem{$ASK_{trend}$}\defterm{Trend Available Seat Kilometers}
\defitem{$LF$}\defterm{Load factor or seat occupancy rate (RPK/ASK)}
\defitem{$ASK$}\defterm{Total Available Seat Kilometers}
\defitem{$ASK_m$}\defterm{Available Seat Kilometers for market $m$}
\defitem{$S_m$}\defterm{Share of trend supply for market $m$}
\defitem{$SR_m$}\defterm{Supply-Shift ratio for market $m$}
\end{deflist}
\end{nomenclature}

\begin{nomenclature}
\textit{Variables (continued)}
\begin{deflist}[AAAAAAA] 
\defitem{$\epsilon_p$}\defterm{Price elasticity of demand}
\defitem{$p_{cap}$}\defterm{Capped ticket price}
\defitem{$p_{trend}$}\defterm{Trend ticket price}
\defitem{$D_{cap}$}\defterm{Capped demand}
\defitem{$D_{trend}$}\defterm{Trend demand}
\defitem{$\theta_{avoidance}$}\defterm{Annual burden associated with demand aversion}
\defitem{$d_f$}\defterm{Social discount rate}
\defitem{$\Theta_{avoidance}$}\defterm{Present valuation of future policy burdens}
\defitem{$S\text{max}_{a}$}\defterm{Maximum market share of aircraft $a$}
\defitem{$t_{ramp}$}\defterm{Ramp duration for fleet replacement}
\defitem{$\tau_{fleet}$}\defterm{Fleet replacement time constant}
\defitem{$ASK_{aircraft}$}\defterm{ASK covered by a specific aircraft type}
\defitem{$EC_{aircraft}$}\defterm{Energy consumption per ASK for an aircraft type}
\defitem{$C_{aircraft}$}\defterm{Total energy consumed by an aircraft type}
\defitem{$C\text{direct}_{\text{energy}}$}\defterm{Direct embarked consumption of each energy carrier}
\defitem{$IF_{\text{pathway}}$}\defterm{Impact factor per unitary production for a pathway}
\defitem{$IF\text{direct}_{\text{pathway}}$}\defterm{Direct impact generated at production for a pathway}
\defitem{$CF_{p, i}$}\defterm{Consumption of input $i$ per unitary production of pathway $p$}
\defitem{$IF_{\text{energy}}$}\defterm{Energy mean impacts weighted by pathway shares}
\defitem{$S_p$}\defterm{Share of production for pathway $p$}
\defitem{$P_{\text{energy}}$}\defterm{Total production of an energy type}
\defitem{$P_{\text{pathway}}$}\defterm{Production of a specific pathway}
\defitem{$C_{\text{pathway, input}}$}\defterm{Input consumption of a pathway}
\defitem{$C_{\text{energy, input}}$}\defterm{Aggregated input consumption by energy type}
\defitem{$C_{\text{resource}}$}\defterm{Total consumption of a resource}
\defitem{$s_{\text{resource}}$}\defterm{Allocation share for a resource}
\defitem{$P_{\text{resource}}$}\defterm{Total production of a resource}
\defitem{$B_{CO_2}$}\defterm{Target carbon budget}
\defitem{$s_{CO_2}$}\defterm{Fair share of carbon budget for aviation}
\defitem{$CO_2$}\defterm{CO2 emissions}
\end{deflist}
\end{nomenclature}

\newpage

\appendix

\section{Models}
\label{sec:models}

This section is dedicated to detailed explanation on modeling assumptions and resulting equations, providing as well a comparison with methods from related works.

\subsection{Time-dependent controls}

First-order delays are commonly used to account for simplified inertia regarding: measuring and reporting information, receiving information and decisions being made, and even for decisions to have a visible effect on the state of a system \cite{meadows_thinking_2009, sterman2000}. While the present work automates decision-making based on system outcomes, we still account for delays regarding the outcomes of time-dependent optimization variables (here called controls). Furthermore, this also allowed for numerical benefits, such as reduced dimensionality and improved the optimization stability. Each output variable $o(t)$ is modeled as an Ordinary Differential Equation (Eq. \ref{eq:delay}), which is dependent on a given input variable $i(t)$ and a delay-time $\tau$. 


\begin{equation}
    \frac{d}{dt}o(t) = \frac{i(t)-o(t)}{\tau}
    \label{eq:delay}
\end{equation}

The supply shift ratio (Eq. \ref{eq:segmented-avoidance}) and the shares of pathway production (Eqs. \ref{eq:impact-energy} and \ref{eq:prod-pathway}) are modeled with inputs that are coarsely discretized in time (2.5 years, while the simulation step is 1 year), here the input values at each time are optimization variables.

The shares of aircraft market penetration are modeled with a ramped pulse, parameterized from 4 optimization variables (ramp start year, max value, lifetime, and ramp-down duration).

\subsection{Air traffic demand}

Many drivers can be linked to the growth in air traffic demand: population, disposable income, trade volumes, fuel prices, urbanization \cite{transport_managing_2003}.

Equation \ref{eq:demand-gcam} presents the model used in the Global Change Analysis Model (GCAM) \cite{kim_gcam_2006, gcam-2019}, where $Pop$ is population, $I$ is per-capita income, $p$ is ticket price, $\epsilon_{\upsilon}$ is the elasticity of demand to variable $\upsilon$, and $\sigma$ is a calibration constant.

\begin{equation}
    D = \sigma Pop\ I^{\epsilon_I} p^{\epsilon_p}
    \label{eq:demand-gcam}
\end{equation}

Yet, there are several issues with using constant elasticities for forecasting aviation demand over long time-horizons. Meta-analyses categorize it as a luxury good and immature market \cite{gallet_elasticity_2014}, and post-COVID studies demonstrate how income elasticities may change rapidly depending on the state of the business cycle (normal, downturn, recovery) \cite{hanson_elasticity_2022}.

In the context of general transportation, these models are also limited to account for demand saturation as personal incomes rise, resulting in ever-growing demand volumes as the Gross Domestic Product (GDP) grows, and that by using S curves (Logistic, Gompertz, or Richards sigmoid functions), to account for the income effect, can produce better estimates for both developed and developing countries \cite{andrieu_modelling_2023}. This method was first applied for estimating personal vehicle stocks from personal income \cite{sommer_vehicle_2007}. Results show that income elasticities can vary significantly as countries develop, but one shortcoming of the model is ignoring the effect of prices in modifying the demand.

In order to account for the phases of growth, maturation and saturation observed in aviation demand \cite{vedantham_aircraft_1994}, a generalized logistic function is used to estimate trend per-capita demand from per-capita income.
In Equation \ref{eq:demand-income}, $L$ and $R$ are the left and right asymptotes (personal propensity to travel at $0$ and $\infty$ personal income), $\iota$ is the income per capita at the inflection point, $B$ is the logistic growth rate, $C$ controls the duration of the transition from $L$ to $R$, and $\nu$ controls near which asymptote maximum growth occurs ($\nu=1$ would yield an logistic equation and $\nu\to0^+$ would tend to a Gompertz function).


\begin{equation}
    RPKpc_{trend}(t)= L + (R-L) \left( \frac{C}{C+\exp(-B(I(t)-\iota))} \right)^{1/\nu}
    \label{eq:demand-income}
\end{equation}

\begin{equation}
    \frac{RPKpc_{story}(t)}{RPKpc_{trend}(t)}= 1 + \frac{F_{SSP}}{1+\exp\left(- \frac{I(t)-2I_{2024}}{2I_{2024}}\right)} 
    \label{eq:demand-story}
\end{equation}

\begin{equation}
    RPK_{trend}(t)=Pop(t)\ RPKpc_{story}(t)
    \label{eq:demand-total}
\end{equation}



\begin{figure*}
	\centering
	\includegraphics[width=.8\textwidth]{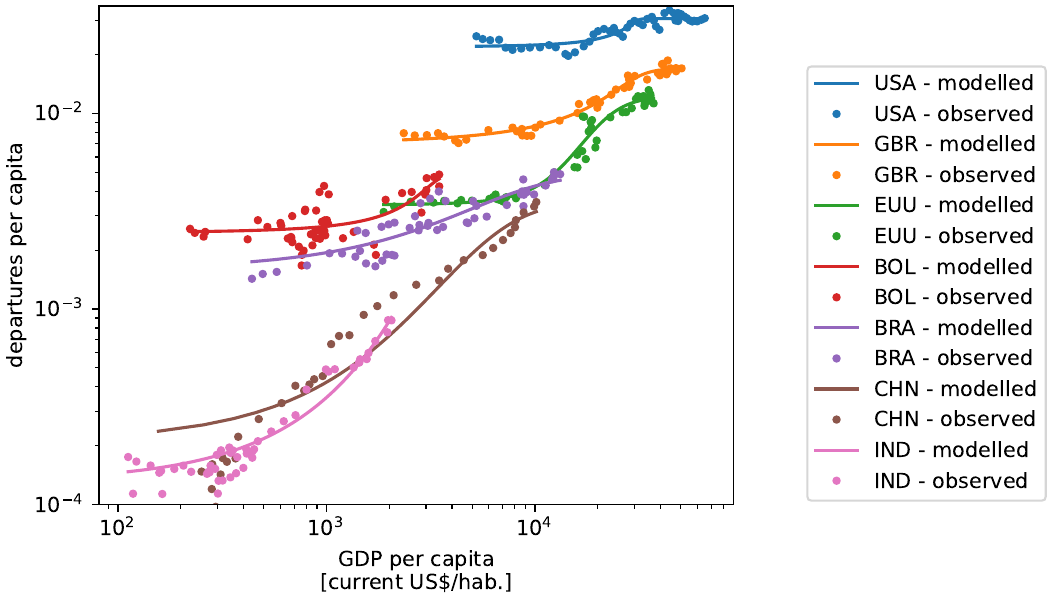}
	\caption{Regional calibration of registered carrier departures per capita as a generalized logistic function of income per capita. Region code: GBR - United Kingdom, USA - United States of America, EUU - European Union, BOL - Bolivia, BRA - Brazil, CHN - China, IND - India. Data from \cite{world_bank}.}
	\label{fig:regional-demand}
\end{figure*}

The parameter set $\theta=(L, R, \iota, B, C, \nu)$ must be calibrated with historical data. In Figure \ref{fig:regional-demand}, World Bank data \cite{world_bank} provided for the demand (carrier departures), population, and income proxies on a regionalized level. But as the traveled distance is highly important in determining total emissions, in the present work, mitigation scenarios are driven by RPK demand, therefore ICAO data was used and calibrated on a global level over the 1980-2019 period. COVID years were excluded from the calibration data, and its after-effects were considered by assuming that by 2024 per capita traffic will reach 2019 levels, this is achieved shifting the parameter $\iota$ by the gap in income per capita between 2024 and 2019.

Besides the logistic trend calibrated based on historical relationships, scenario narratives are further endogeneized by accounting a storyline multiplier factor to trend demand (Eq. \ref{eq:demand-story}), which is essentially another S-curve that goes from 1 to $F_{SSP}$ with inflection at around 1.5 times 2024 income per capita. For SSP scenarios 2, 3, and 4 $F_{SSP}=1$, meaning they will follow the historically calibrated logistic function. For scenario 1, $F_{SSP}=0.9$ stabilizing demand 90 \% lower than trend; while for scenario 5 $F_{SSP}=1.5$.

The supply, in terms of ASK, is then estimated using Load Factor (Eq. \ref{eq:load-factor}) that grows following a quadratic curve in time \cite{planes_simulation_2021} from 82.4 \% in 2019 to 92 \% in 2075.

\begin{equation}
    LF(t) = RPK_{trend}(t) / ASK_{trend}(t)
    \label{eq:load-factor}
\end{equation}







\subsection{Aircraft design}
\label{subsec:aircraft-design}

Modifying the energy carrier changes and the propulsion system architecture requires specific technology, e.g., cryogenic fuel tank, fuel cells, electric motors, all of which are expected to mature at different rates. To demonstrate this, several sources are used to provide technology parameters and expected year of entry-into-service, which include research papers on green transportation technologies \cite{wallington_green_2024, adler_energy_2025} technology roadmaps from IATA \cite{iata_tech_2050} and ATI \cite{ati_aerodynamic, ati_cryogenic, ati_electrical, ati_fuelcell}, ICCT aircraft design studies \cite{icct_hydrogen, icct_electric, icct_fuelcell}  NASA electric propulsion studies and technology aspiration \cite{felder_nasa_2015, papathakis_nasa_2017, woodworth_nasas_nodate, bradley_subsonic_2015} and EASA type certificate data \cite{easa234}. Then, conservative-to-optimistic technology scenarios are made by interpolating the higher or lower bound of parameters in time.

\begin{table*}[width=.75\textwidth,cols=6,pos=h]
    \caption{Quantitative evolution of aircraft technology parameters.}
    \centering
    \begin{tabular*}{\tblwidth}{c|c|c c c|c}
    \toprule
    Technology Parameter & Unit  & 2020 & 2040    & 2060     & Sources   \\
    \midrule
    Battery Specific Energy     & Wh/kg & 200  & 350-800 & 600-1500 & \cite{iata_tech_2050, icct_electric, felder_nasa_2015, bradley_subsonic_2015, woodworth_nasas_nodate} \\
    E-motor Specific Power      & kW/kg & 2    & 10-25   & 15-28 & \cite{ati_electrical, papathakis_nasa_2017} \\
    Electronics Specific Power  & kW/kg & 2    & 15-25   & 20-32 & \cite{ati_electrical, woodworth_nasas_nodate} \\
    Fuel cell Specific Power    & kW/kg & 1    & 2-3     & 3-6 & \cite{iata_tech_2050, ati_fuelcell, papathakis_nasa_2017, wallington_green_2024} \\
    Fuel cell Efficiency        & \%    & 40  & 45-55   & 50-65 & \cite{ati_fuelcell, wallington_green_2024} \\
    LH2 tank Gravimetric Index  & \%    & 20   & 30-65   & 35-80 & \cite{iata_tech_2050, ati_cryogenic, icct_hydrogen, icct_fuelcell, wallington_green_2024} \\
    Structural weight reduction & \%    & 0    & 15-30   & 20-40 & \cite{iata_tech_2050, ati_aerodynamic} \\
    \bottomrule
    \end{tabular*}
    \label{tab:aircraft-tech}
\end{table*}

\begin{figure*}
    \centering
    \includegraphics[width=0.8\textwidth]{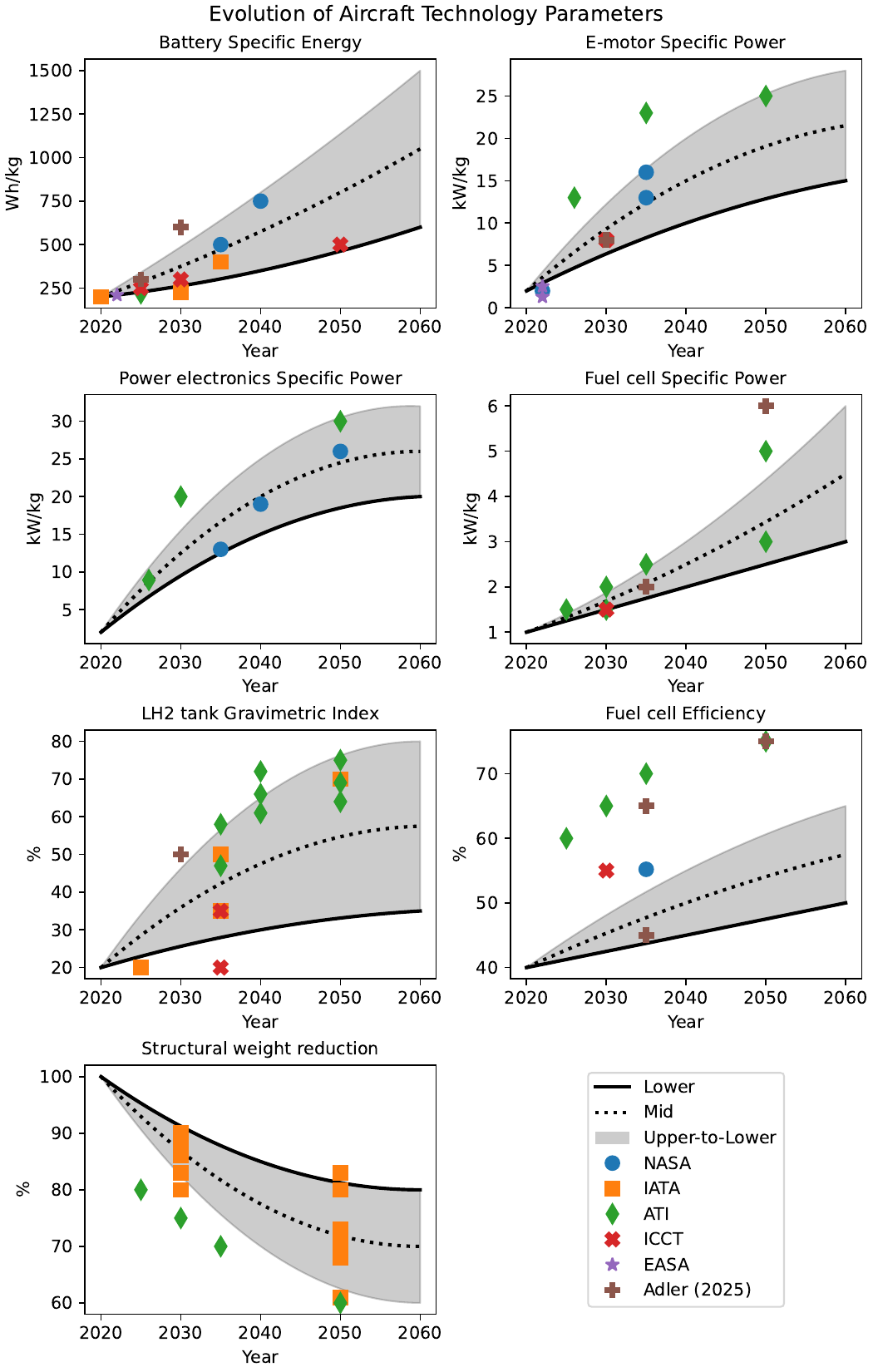}
    \caption{Improvement in selected aircraft parameters as a function of expected Entry-Into-Service. Filled between the upper and lower limit for the technology scenarios, solid line shows lower technology scenario, and dotted line shows mid technology scenario.}
    \label{fig:aircraft-tech}
\end{figure*}

\begin{table}[width=\columnwidth,cols=3,pos=h]
    \caption{TLAR determined by market segment.}
    \centering
    \begin{tabular}{c|c c c}
    \toprule
    Category & Range (km)  & Seats & Lifetime\\
    \midrule
    General & 500 & 19 & 18.5\\
    Commuter & 1500 & 50 & 15.3\\
    Regional & 4500 & 80 & 23.5\\
    Short-medium & 8000 & 120 & 26.6\\
    Long range & 15000 & 250 & 24.6\\
    \bottomrule
    \end{tabular}
    \label{tab:tlar-category}
\end{table}

\begin{table}[width=\columnwidth,cols=3,pos=h]
    \caption{TLAR determined by aircraft architecture.}
    \centering
    \begin{tabular}{c|c c}
    \toprule
    \multirow{2}{*}{Architecture} & Speed & Altitude\\
    & (Mach) &  (thousand ft)\\
    \midrule
    Jet-A and Gas Turbine & 0.75 & 27 \\
    LH2 and Gas Turbine & 0.75 & 27 \\
    Battery and E-motor & 0.5 & 20 \\
    LH2 fuel-cell and E-motor & 0.5 & 20 \\
    
    \bottomrule
    \end{tabular}
    \label{tab:tlar-architecture}
\end{table}

Table \ref{tab:aircraft-tech} present the lower and upper values used for the technology parameters, according to entry-into-service (EIS) and Figure \ref{fig:aircraft-tech} displays how some key parameters are interpolated in time and how they compare with the sources used. The main goal of this is to be able to account for the trade-off regarding the timing of deployment of aircraft architectures: early deployment of maturing technology and lock-in with mediocre performance, or late deployment with mature and better performances.

In order to avoid over-reliance on optimist technology, especially in a sector that has missed many of its recent environmental targets \cite{missed-targets}, some sources that are purposefully left out of the technology scenario range. This is of particular importance for fuel-cell systems because:

\begin{itemize}
    \item Fuel-cells have limited power output, so several cells have to be stacked to compose total power, decreasing power-to-weight ratio of the system;
    \item Narrow ranges of operating temperature plus high thermal losses, means these systems require thermal management for high-power applications, adding further weight (which is included in the fuel-cell specific power);
    \item Aircraft-tailored applications needs much higher power output relative to automotive fuel cells, overall studies may display diverging assumptions on whether the larger scale will result in improved \cite{icct_fuelcell} or degraded efficiency performance \cite{seitz_initial_2022}.
\end{itemize}

The Top-Level Aircraft Requirements (TLAR) are split into two sets: the number of seats and range are determined solely by the market segment (Table \ref{tab:tlar-category}), the cruise speed and altitude are determined by the propulsion architecture (Table \ref{tab:tlar-architecture}). Overall, gas turbines can yield efficient operation at higher and faster flight conditions relative to propellers, but their cruise altitude and speed were purposefully limited in the present work, based on recent findings that re-designing aircraft to fly lower and slower can significantly reduce non-CO2 impacts at less than 1 \% extra operating cost \cite[][Figures 6.5, 6.8, 6.13, and Table E.4]{proesmans_thesis_2024}.

The Generic Airplane Model (GAM) \cite{kambiri_energy_2024}, from ENAC, is then used as a preliminary airplane design tool. It uses regression of historical airplane data to estimate airframe and structural weight and adds the propulsion system mass depending on the technology components that each architecture uses.

\subsection{Fleet deployment}

The global aircraft fleet is segmented into distance bands, rather than regionally or by routes. The 2019 repartition of ASK and emissions obtained from the AeroSCOPE database \cite{salgas_aeroscope} are also used to compare current energy consumption to that of new aircraft designs.

\begin{equation}
\label{eq:segmented-avoidance}
\begin{split}
    ASK(t) & = \sum_{m\ in\ \text{markets}} ASK_m(t)\\
    & =\sum_{m\ in\ \text{markets}} S_{m}\ ASK_{trend}(t)\ (1 - SR_{m} (t) )
\end{split}
\end{equation}

Each market is assigned a constant share of trend supply, and subject to a demand-side policy that avoids part of the trend demand (Eq. \ref{eq:segmented-avoidance}). The Supply-Shift Ratio $SR$ is treated as a time-dependent control, whose input values are used as optimization variables in the low-demand formulation, and it represents the part of trend supply that is to be avoided ($0$: all trend supply is met, $1$: all flights are banned).

\begin{equation}
    \label{eq:elast_price}
    \frac{dp}{p}=\frac{1}{\epsilon_p}\frac{dD}{D} \implies \ln\left(\frac{p_{cap}}{p_{trend}}\right) = \frac{1}{\epsilon_p} \ln\left(\frac{D_{cap}}{D_{trend}}\right)
\end{equation}

In Equation \ref{eq:burden-time}, the annual burden associated to demand aversion is defined, formulated as the relative ticket price increase due to the chosen value for $SR$. Under the assumptions that: demand reduction ultimately increase consumer prices (either caused by a tax or as a consequence of scarce supply), price elasticity $\epsilon_P$ is constant in time and among markets. From the definition of price elasticity $\epsilon_p=\frac{dRPK/RPK}{dp/p}$, one can derive by integration (Eq. \ref{eq:elast_price}) that the relative ticket price change is equal to $(1-SR)^{1/\epsilon_p}$ to the relative demand change. While the assumptions are not applicable to reality due to market-specific elasticities \cite{iata-elasticities}, they may still serve as simplified way to compare the suitability of scenarios with demand-aversion strategies without recurring to cost estimations.

In Equation \ref{eq:burden-avoidance}, the objective function of the low-demand formulation is defined as the present valuation of future policy burdens. It is a time-integration of the annual burden of demand avoidance, multiplied by a discount factor. The social discount rate $d_f$ is the rate at which future burdens are undermined relative to present burdens, which was set at 3 \%. There is, however, a significant debate \cite{goulder_choice_2012, schoenmaker_which_2024} on whether this parameter should be defined by normative (how policies should be put in place) or positive (how policies likely will be put in place) approaches, the value chosen stands as a middle ground between the normative and the positive range.

\begin{equation}
\label{eq:burden-time}
    \theta_{avoidance}(t) = \frac{\Delta P}{P} = \frac{\sum_{m\ in\ \text{markets}} S_m \left(1-SR_m(t)\right)^{1/\epsilon_p}}{ASK(t)}
\end{equation}

\begin{equation}
\label{eq:burden-avoidance}
\begin{split}
    \Theta_{avoidance} = & \int_{t_0}^{t_1} (1+ d_f)^{t_0-t}\ \theta_{avoidance}(t)\ dt
\end{split}
\end{equation}

The market share of each aircraft type is also modeled using time-dependent controls (Eq. \ref{eq:delay}), but subject to a parameterized ramped pulse shape. The input signal is treated as zero before the year of Entry-Into-Service EIS, and then grows from 0 to a max share $S\text{max}$ linearly for a duration of $t_{ramp}$. EIS and $S\text{max}$ are tailored for each aircraft design and used as optimization variables. $t_{ramp}$ is parameterized as $2\tau_{fleet}$ to maintain the ramped step shape. and $\tau_{fleet}$ is kept as 4 years, meaning it takes 20 years from the introduction of a new aircraft to replace 98 \% of the fleet \cite{delbecq-fleet}.

The total energy carrier consumed by aircraft operations is estimated from covered supply and design energy consumption (sub-section \ref{subsec:aircraft-design}) as shown in Equation \ref{eq:aircraft-consumption}. Then, the direct consumption of each energy carrier is aggregated as the sum among the carrier-consuming aircraft in Equation \ref{eq:aircraft-aggregation}.

\begin{equation}
    C_{aircraft} (t) = ASK_{aircraft}(t) EC_{aircraft}
    \label{eq:aircraft-consumption}
\end{equation}

\begin{equation}
    C\text{direct}_{\text{energy}} = \sum_{a\ in\ \text{energy architectures}} C_{a}
    \label{eq:aircraft-aggregation}
\end{equation}



\subsection{Energy-Mix}

This work considers different energy carriers, and for some of them several production pathways are available, such as synthetic fuels. Moreover, some energy carriers can be consumed in the production of other carriers and some of them can compete for the same primary energy sources or materials. In that regards, it is necessary to consider a modular implementation to the energy production model.

Each energy conversion process is modeled in a modular manner as a production pathway. Pathways have a specific consumption of input flows per unitary production of output flows. The energy mix assembles all the production processes and the links them to calculate both intensive and extensive properties of the energy production system.

The computation of the impact generated (only CO2 emissions in this work, but any cumulative impact could be extended, such as land required, water consumption, ...) per unitary production, an intensive quantity, is made from primary-to-final. This is mainly due to the fact that the impacts made in the production of inputs must be known beforehand in order to be accounted for in the indirect impacts of outputs. On the other hand, the aggregated consumption and production of energies, an extensive quantity, is made from final-to-primary, because total consumption of final energies determine the required consumption of inputs, which determines how much production of each energy-input is required.

Consider the energy system required to produce two final energy carriers: Jet-A and liquid hydrogen. Jet-A is produced from a mix of fossil kerosene, and synthetic kerosene (electrofuel). Gaseous hydrogen is produced from electrolysis and is used to produce both liquid hydrogen, through liquefaction, and electrofuel, through power-to-liquid pathways.
The emission factor ($EF$) of electrofuel and liquid hydrogen are both dependent on the $EF$ of gaseous hydrogen, which depends on that of electricity. The computation of $EF$ starts with electricity (primary), then gaseous hydrogen (secondary), then liquid hydrogen (final) and electrofuel (secondary), then Jet-A (final). The total production of each energy, however, must follow the inverse path: from final-to-primary. Taking the same example: Power-to-liquid production depends on Jet-A consumed, and liquefaction production depends on liquid hydrogen consumed. Electrolysis production depends on the gaseous hydrogen consumed in liquefaction and power-to-liquid production. Finally, electricity production required is estimated from consumption in the production of gaseous hydrogen. The computation of aggregated production and consumption starts with Jet-A (final) and liquid hydrogen (final), then electrofuel (secondary), then gaseous hydrogen (secondary), then electricity (primary).

In some implementations of modular energy system models, the estimation of properties is made altogether (intensive and extensive) per energy type and pathway \cite{witness}. This creates a coupled model: intensive quantities depend on the mix of the upstream system and extensive quantities depend on the aggregated consumption of the downstream system. This yields that an initial guess must be made up- and downstream, which is then iteratively solved until convergence. By separating modules responsible for estimating intensive and extensive quantities, if the system has no retroaction, the coupling disappears and direct computation can be achieved.

\subsubsection{Intensive impacts}

\begin{table*}[width=.92\textwidth,cols=8,pos=h]
    \caption{Energy consumption and direct emissions for each of the production pathways considered. For pathways under technology maturing, the two values represent the 2025 and 2050 values.}
    \centering
    \begin{tabular*}{\tblwidth}{c c|c c c c|c|c}
    \toprule
    \multirow{2}{*}{Energy Carrier} & \multirow{2}{*}{Production Pathway}  & \multicolumn{4}{c|}{Energy consumption (MJ/MJ)} & Direct emissions  & \multirow{2}{*}{Source}   \\
    & & Oil & Biomass & Electricity & Gas H2 & (g CO2 / MJ) & \\
    \midrule
    Fossil Jet-A & Refinery & 1.16 & - & - & - & 15.5  &  \cite{jing_understanding_2022, stratton_impact_2011} \\
    \hline
    \multirow{3}{*}{Biofuel} & HEFA & - & 1.95 & - & - & 62.73 & \multirow{3}{*}{\cite{NEULING201854}} \\
    & ATJ & - & 3.33 & - & - & 51.55 & \\
    & FT & - & 5.0 & - & - & 35.3 & \\
    \hline
    \multirow{2}{*}{Gas H2} & Gas reforming & - & - & - & - & 101.5 & \cite{ji_h2} \\
    \cline{8-0}
     & Electrolysis & - & - & 1.41-1.33 & - & 0 & \multirow{3}{*}{\cite{wallington_green_2024}} \\
    \cline{1-7}
    Electrofuel & Power-to-liquid & - & - & 0.65-0.56 & 1.89-1.68; & 0  &   \\
    \cline{1-7}
    Liquid H2 & Liquefaction &  - & - & 0.22-0.16 & 1.0 & 0 &  \\
    \bottomrule
    \end{tabular*}
    \label{tab:energy-prod}
\end{table*}

\begin{table*}[width=.8\textwidth,cols=4,pos=h]
    \caption{Emission factor associated to energy inputs and brief description of the method to estimate them. For pathways under technology maturing, the two values represent the 2025 and 2050 values.}
    \centering
    \begin{tabular*}{\tblwidth}{c|c|c c}
    \toprule
    \multirow{2}{*}{Energy input} & Emission factor & \multirow{2}{*}{Method} & \multirow{2}{*}{Source}\\
     & (g CO2 / MJ) & & \\
    \midrule
    Oil & 63.32 & Kerosene combustion divided by oil consumption & \cite{stratton_impact_2011}\\
    Biomass & 0 & Accounted only as biofuel direct emissions & \cite{NEULING201854}\\
    Electricity & Scenario-dependent & Final consumption divided by electricity emissions & \cite{ar6_database}\\
    \bottomrule
    \end{tabular*}
    \label{tab:energy-input}
\end{table*}

Impacts generated in the production of energy carriers are heavily dependent on the efficiency of processes and the impacts of consumed inputs, and these are mainly determined by the background energy system \cite{mendoza_beltran_when_2020}. Recent works have highligthed the importance of linking global scenarios to perform prospective life-cycle assessment of energy \cite{sacchi_premise_2022}, and have also been applied for the prospective assessment of climate neutral aviation \cite{sacchi_how_2023}, showing a great increase in emissions associated with the production of synthetic jet fuel when a 3.5°C temperature increase scenario is chosen instead of a 2°C one.

The impact factor $IF_{\text{pathway}}$ (Eq. \ref{eq:impact-pathway}), impact per unitary production, of each output flow associated to production pathways is modeled as the sum of direct impact generated at production plus the impacts associated to the input flows consumed in the process. $CF_{p, i}$ is the consumption of input $i$ per unitary production of pathway $p$. This is made because the inputs consumed in the process have impacts themselves, and by consuming them these indirect impacts must be accounted in the impacts of the output flow.

Table \ref{tab:energy-prod} summarizes the production pathways for each of the accounted energy carriers, their energy consumption and direct emissions per produced output. Technology maturing was accounted for some production pathways with an inverse consumption (analogous to process efficiency) that decreases linearly until stagnation in 2050. Table \ref{tab:energy-input} summarizes the emission factor associated to the consumption of each energy input, such than total emissions are a sum of direct emissions and indirect emissions, and Figure \ref{fig:electricity-emission} shows the electricity emission factor for electricity for a set of SSP background scenarios.


\begin{equation}
\label{eq:impact-pathway}
\begin{split}
    IF_{\text{pathway}}= & IF\text{direct}_{\text{pathway}}\\
    &+\sum_{i\ in\ \text{pathway inputs}}CF_{\text{pathway}, i}\ IF_i
\end{split}
\end{equation}

Because each energy type can be produced by several pathways, a mixing process is applied where the energy mean impacts $IF_{\text{energy}}$ (Eq. \ref{eq:impact-energy}) is weighted by the share of pathway production, which are treated as a time-dependent control, used as optimization variables.

\begin{equation}
    IF_{\text{energy}}=\sum_{p\ in\ \text{energy pathways}}S_p\ IF_p
    \label{eq:impact-energy}
\end{equation}

\subsubsection{Extensive production and consumption}

The production of each energy type (Eq. \ref{eq:prod-energy}) is the direct energy consumption (directly embarked in aircraft) plus what was consumed to make other energy types. If $e_0$ is an energy input to $e_1$, $e_1$ is an energy output to $e_0$. The computation is initialized with final energies because the term $\sum_{o}(CF_{o, e}\ P_o)$ is zero, as no intermediate processes consume them.

\begin{equation}
    P_{\text{energy}}=C\text{direct}_{\text{energy}} + \sum_{o\ in\ \text{energy outputs}}CF_{o, \text{energy}}\ P_o
    \label{eq:prod-energy}
\end{equation}

\begin{equation}
    P_{\text{pathway}}=S_{\text{pathway}}\ P_{\text{energy}}
    \label{eq:prod-pathway}
\end{equation}

\begin{equation}
    C_{\text{pathway}, \text{input}}=CF_{\text{pathway}, \text{input}}\ P_{\text{pathway}}
    \label{eq:cons-pathway}
\end{equation}

\begin{equation}
    C_{\text{energy}, \text{input}}=\sum_{p\ in\ \text{pathways}} C_{p, \text{input}}
    \label{eq:cons-energy}
\end{equation}

The production of each pathway (Eq. \ref{eq:prod-pathway}) is then estimated from pathway share. Input consumption of pathways are estimated (Eq. \ref{eq:cons-pathway}) and then are aggregated by energy type (Eq. \ref{eq:cons-energy}). The process is then repeated for each energy type until primary energies.

\subsubsection{Consumption and impacts constraints}

For biomass and electricity, the total consumption is constrained applying the concept of an allocation principle, initially developed for Absolute Environmental Sustainability Assessments \cite{bjorn_framework_2019, hjalsted_sharing_2021}, but applied for energy production (Equation \ref{eq:conso_constraint}). Because there is little consensus on how to find such fair shares \cite{pais_current_2024}, two different values were explored: one reflecting a conservative energy availability (5.0 \%), and another reflecting a preferential availability to the aviation sector (8.6 \%). Both values use some sort of grandfathering, which tend to lock the economic system into its present state. Yet, values are still conservative when compared to other institutional roadmaps \cite{becken_implications_2023}.

The conservative value was obtained based on a reference mitigation scenario, the IMP-REN-2.0, in which 38.6 \% of the biomass production is allocated to the transport sector \cite[Figure 6.1]{energy_ar6_wg3}, the fair share allocated to aviation is considered to be 13 \% of that, which is the 2019 sector's share of oil consumption relative to the entire transportation consumption \cite{IEA2019oil}. The preferential access value was obtained based on the sector's 2019 share of global oil consumption \cite{IEA2019oil}.

\begin{equation}
    \label{eq:conso_constraint}
    C_{\text{resource}} \le s_{\text{resource}} P_{\text{resource}}
\end{equation}

In the low demand formulation, the cumulative emissions of the sector is a constraint rather than the objective to minimize (Eq. \ref{eq:co2_constraint}). In these cases, we assumed the target carbon budget $B_{CO_2}$ to be the remaining 2°C carbon budget with 66 \% confidence \cite{lamboll_assessing_2023}, and the fair share to be 3.0 \%, which is the sector's share of direct, indirect and induced GDP \cite{atag_benefits}.

\begin{equation}
    \label{eq:co2_constraint}
    \int_{t_0}^{t_1} CO_2(t) dt \le s_{CO_2} B_{CO_2}
\end{equation}

\bibliographystyle{elsarticle-num}

\bibliography{bibliography}





\end{document}